\documentclass[fleqn,usenatbib]{mnras}

\usepackage[T1]{fontenc}
\usepackage{ae,aecompl}

\usepackage{graphicx}	
\usepackage{amsmath}	
\usepackage{amssymb}	
\usepackage{amsfonts}
\usepackage{hyperref}
\usepackage{cleveref}
\crefname{section}{§}{§§}
\Crefname{section}{§}{§§}
\usepackage{longtable}
\usepackage{natbib}
\usepackage{verbatim}
\usepackage{lscape}
\usepackage{graphics}
\usepackage{epsfig}
\usepackage{multirow}
\usepackage{setspace}
\usepackage{bigdelim}
\usepackage{bigstrut}
\usepackage{floatflt}	
\usepackage{wrapfig}
\usepackage{multicol}

\usepackage{indentfirst}

\usepackage{caption}
\usepackage{subcaption}
\usepackage{verbatim}  
\usepackage{float}
\usepackage{xcolor}     
\usepackage{siunitx}
\usepackage{rotating}

\sffamily
\def\h1{H\,{\sc i}}

\def\ga{\raisebox{-0.3ex}{\mbox{$\stackrel{>}{_\sim} \,$}}}
\def\la{\raisebox{-0.3ex}{\mbox{$\stackrel{<}{_\sim} \,$}}}

\def\c1{C\,{\sc i}}

\def\NH3{NH$_{3}$}
\def\ch3cn{CH$_{3}$CN}

\def\deg{$^{o}$}

\def\kms{km s$^{-1}$}
\def\apj{ApJ}
\def\apjs{ApJSS}
\def\apjl{ApJL}
\def\aa{A\&A}
\def\aap{A\&A}

\def\aj{AJ}

\def\mnras{MNRAS}

\defcitealias{obreschkow14}{OG14}
\defcitealias{obreschkow16}{O16}

\title[The influence of AM and environment on the HI gas of late-type galaxies]{
The influence of angular momentum and environment on the \h1 gas of late-type galaxies 
}
\author[C. Murugeshan et al.]{Chandrashekar Murugeshan,$^{1,2}$\thanks{E-mail:cmurugeshan@swin.edu.au}
Virginia Kilborn,$^{1,2}$ Thomas Jarrett,$^{3}$ \newauthor O. Ivy Wong,$^{4,5,2}$ Danail Obreschkow,$^{5,2}$ Karl Glazebrook,$^{1,2}$ \newauthor
Michelle E. Cluver,$^{1,6}$ and Christopher J. Fluke$^{1}$ \\
$^{1}$Centre for Astrophysics and Supercomputing, Swinburne University of Technology, Hawthorn, Victoria 3122, Australia \\
$^{2}$ARC Centre of Excellence for All Sky Astrophysics in 3 Dimensions (ASTRO 3D), Australia \\
$^{3}$Astronomy Department, University of Cape Town, Private Bag X3, Rondebosch 7701, Republic of South Africa \\
$^{4}$CSIRO Astronomy \& Space Science, PO Box 1130, Bentley, WA 6102, Australia \\
$^{5}$ICRAR-M468, UWA, 35 Stirling Highway, Crawley, WA 6009, Australia \\
$^{6}$Department of Physics and Astronomy, University of the Western Cape,Robert Sobukwe Road, Bellville, South Africa}

\date{Accepted XXX. Received YYY; in original form ZZZ}

\pubyear{2018}
\begin{document}
\label{firstpage}
\pagerange{\pageref{firstpage}--\pageref{lastpage}}
\maketitle
\begin{abstract}
We use high-resolution \h1 data from the WHISP survey to study the \h1 and angular momentum properties of a sample of 114 late-type galaxies. We explore the specific baryonic angular momentum -- baryonic mass ($j_b - M_b$) relation, and find that an unbroken power law of the form $j_b \propto M_b^{0.55 \pm 0.02}$ fits the data well, with an intrinsic scatter of $\sim 0.13 \pm 0.01$ dex. We revisit the relation between the atomic gas fraction, $f_{atm}$, and the integrated atomic stability parameter $q$ (the $f_{atm} - q$ relation), originally introduced by Obreschkow et al., and probe this parameter space by populating it with galaxies from different environments, in order to study the influence of the environment on their $j_b$, $f_{atm}$ and $q$ values. We find evidence that galaxies with close neighbours show a larger intrinsic scatter about the $f_{atm} - q$ relation compared to galaxies without close-neighbours. We also find enhanced SFR among the deviating galaxies with close neighbours. In addition, we use the bulge-to-total (B/T) ratio as a morphology proxy, and find a general trend of decreasing B/T values with increasing disc stability and \h1 fraction in the $f_{atm} - q$ plane, indicating a fundamental link between mass, specific angular momentum, gas fraction and morphology of galaxies. 
\end{abstract}

\section{Introduction}
\label{Sec:Intro}

Galaxy evolution is governed by a range of internal and external processes. The former include fundamental properties such as mass and angular momentum, in conjunction with non-axisymmetric potentials and feedback processes (see for example Jog~\citeyear{jog02}; Zasov \& Zaitseva~\citeyear{zasov17}; Genel et al.~\citeyear{genel15}). External processes are particularly prevalent in higher density environments such as compact groups and clusters, where gas and stars can be redistributed in galaxies due to tidal interactions and ram pressure stripping (Gunn \& Gott~\citeyear{gunn72}; Fasano et al.~\citeyear{fasano2000}). This has been shown to affect global properties of galaxies such as their gas fractions (Davies \& Lewis~\citeyear{davies73}; Giovanelli \& Haynes~\citeyear{giovanelli85}; Solanes et al.~\citeyear{solanes01}), star formation rate (see for example Lewis et al.~\citeyear{lewis02}; G{\'o}mez et al.~\citeyear{gomez03}), colour and morphology (Skibba et al.~\citeyear{skibba09}). In addition, there is a morphological trend with respect to the environment, wherein the fraction of redder and more early-type elliptical galaxies is observed to be higher in dense environments such as clusters compared to the field, where the fraction of disc-like star forming late-type galaxies is higher (morphology--density relation; Dressler~\citeyear{dressler80}; Goto et al.~\citeyear{Goto03}). Therefore, studying the effects of the various processes on the global properties of galaxies is important for our understanding of how galaxies evolve. A particular focus of this work is to study the influence of angular momentum (an intrinsic property) and that of the environment on the observable properties of galaxies.

One of the key constituents of galaxies is their neutral atomic hydrogen (\h1) gas. It is the \h1 gas that is consequently converted to stars via a H$_2$ phase and drives the evolution of galaxies. The \h1 gas disc is loosely bound to the gravitational potential of the galaxy and typically more extended than the stellar disc ($R_{\textrm{\h1}} \sim 2-3 R_{d}$, where $R_{d}$ is the optical disc scale-length, see for example Broeils \& van Woerden~\citeyear{broeils94}; Broeils \& Rhee~\citeyear{broeils97}; Verheijen \& Sancisi~\citeyear{verheijen01}), which leaves the \h1 gas susceptible to environmental processes (Hibbard \& van Gorkom~\citeyear{hibbard96}). In addition, the \h1 gas is an excellent tracer of the rotation velocity of galaxies out to large radii, thus enabling us to accurately compute their angular momentum. High-resolution \h1 observations of galaxies, therefore, prove to be indispensable, as \h1 is a very good tracer of the effects of both internal and external processes.

Peebles (\citeyear{peebles69}) suggested that interacting dark matter (DM) halos acquire their angular momentum (AM) from tidal torques during the protogalactic stages. The baryonic material that is eventually accreted by the DM halos then cools onto a centrifugally supported disc. This leads to the formation of discs with well-defined scale lengths (see Fall \& Efstathiou~\citeyear{fall80}; Mo et al.~\citeyear{Mo98} and references therein), bringing about the various disc scaling relations such as the Tully-Fisher relation (Tully \& Fisher~\citeyear{tully77}), the mass-size relation (Verheijen \& Sancisi~\citeyear{verheijen01}), the fundamental plane of spiral galaxies (Shen et al.~\citeyear{shen02}) and various other \h1 scaling relations (Haynes \& Giovanelli~\citeyear{haynes84}; Chamaraux et al.~\citeyear{chamaraux86}; Solanes et al.~\citeyear{solanes96}; Verheijen \& Sancisi~\citeyear{verheijen01}). Thus, AM along with mass becomes a fundamental underpinning property of galaxies. 

The first empirical study of galactic stellar AM was conducted by Fall (\citeyear{fall83}) who found a tight correlation between the stellar mass ($M_{\star}$) and the specific stellar AM ($j_{\star}$) of the form $j_{\star} = qM^{\alpha}_{\star}$, with $\alpha \approx 2/3$ for both spiral and elliptical galaxies, but with the factor $q$ about five times less for ellipticals. The $\Lambda$CDM model of the universe predicts this relationship between mass and specific AM with an exponent $\alpha = 2/3$. Romanowsky \& Fall (\citeyear{romanowsky12}) revisited this relation with a larger sample of spiral and elliptical galaxies and established the fact that, indeed the morphology of galaxies is related to their sAM. This gives a more physically motivated explanation to the observed range of galaxies in the Hubble classification, in the sense that AM determines the morphology of galaxies (Sandage et al.~\citeyear{sandage70}; Hernandez \& Cervantes-Sodi~\citeyear{hernandez06};). Similarly, a fundamental relation between mass, specific baryonic (cold gas + warm gas + stars) AM and the bulge mass fraction ($\beta$) was discovered for late-type galaxies by Obreschkow \& Glazebrook (\citeyear{obreschkow14})[hereafter \citetalias{obreschkow14}]. Following this, the colour and morphology of galaxies were also observed to be linked to their AM (see for example Cortese et al.~\citeyear{cortese16}; Sweet et al.~\citeyear{sweet18}).

It is therefore important to understand both the evolution of AM in galaxies and the various processes that affect it. With the advancement in semi-analytic and hyrodynamical simulations over the past few decades, studies focusing specifically on the AM evolution of galaxies find that AM in galaxies can be lost due to mergers (Hernquist \& Mihos~\citeyear{Hernquist95}; Lagos et al.~\citeyear{lagos17}), while it can be increased due to cold-mode accretion (see for example Danovich et al.~\citeyear{danovich15}),  galactic winds and fountains (Brook et al.~\citeyear{Brook12}; DeFelippis et al.~\citeyear{DeFelippis17} and references therein). 

In terms of the connection between AM and the \h1 properties of galaxies, Zasov \& Rubtsova (\citeyear{zasov89}) found the first empirical evidence that the \h1 mass of isolated disc galaxies strongly correlates with their sAM. In a similar vain, Huang et al. (\citeyear{huang12}) find that galaxies with higher \h1 gas fractions reside preferentially in dark matter halos with high spin parameters. 

Star formation in disc galaxies is induced via disc instabilities that allow the \h1 gas to collapse to form molecular clouds, where eventually stars are formed. The local disc stability is often quantified by the Toomre parameter 
\begin{equation*}
    Q \approx \frac{\sigma \kappa}{ \pi G \Sigma}
\end{equation*}
Where $\sigma$ is the dispersion velocity of the gas in the disc, $\kappa$ is the epicyclic frequency and $\Sigma$ is the gas surface density (Toomre~\citeyear{toomre64}). A value of $Q < 1$ implies that the gas disc is unstable, promoting star formation and if $Q > 1$, the disc is said to be stable, restricting star formation. Many previous theoretical and simulation studies showed the link between disc stability and AM, where unstable disc galaxies were found to re-distribute their AM and transform into spheroidal systems (Combes et al.~\citeyear{combes90}; Norman, Sellwood \& Hasan~\citeyear{norman96}; Mao \& Mo~\citeyear{mao98}; Dutton \& van den Bosch~\citeyear{dutton12}; Stevens et al.~\citeyear{stevens16a}). 

Obreschkow et al. (\citeyear{obreschkow16})[hereafter \citetalias{obreschkow16}] linked the atomic disc stability of galaxies to their sAM by introducing a parameter-free model predicting a correlation between the atomic gas fraction $f_{atm} = 1.35M_{\textrm{\h1}}/M_{b}$ and what they originally termed the ``global stability" parameter $q = j_{b}\sigma/G M{_b}$, for axisymmetric disc galaxies in equilibrium. Here $M_{\textrm{\h1}}$ is the \h1 mass of the galaxy, $j_b$ is the specific baryonic angular momentum, $\sigma$ is the dispersion velocity of the Warm Neutral Medium (WNM), $M_b$ is the total baryonic mass  and $G$ is the universal gravitational constant. They find that a sub-sample of late-type galaxies in THINGS (Walter et al.~\citeyear{things08}), dwarf galaxies part of LITTLE THINGS (Hunter et al.~\citeyear{littlethings12}), and a sub-sample of confusion-free HIPASS (Meyer et al.~\citeyear{hipass04}) sources follow the model predictions consistently. It is worth noting, a priori, that the $q$ parameter does not refer to the stability of any global mode, but to a local mode (Toomre instability of the atomic gas) that is integrated over the entire disc. For this reason we will term this parameter as the ``integrated atomic stability parameter" (hereafter simply the stability parameter), to avoid any confusion. Furthermore, $q$ does not describe the actual current stability of the disc or a sub-component, but the hypothetical mean stability of a purely atomic disc with the same spin.

Lutz et al. (\citeyear{lutz17}, \citeyear{lutz18}) show that a sample of isolated \h1-excess galaxies follow the $f_{atm} - q$ relation consistently and owe their excess \h1 gas fractions to higher sAM. In a subsequent study testing the analytical model of \citetalias{obreschkow16}, Murugeshan et al. (\citeyear{murugeshan2019}) confirm that AM regulates the \h1 gas fraction in \h1-deficient spirals from low-density environments. D{\v{z}}ud{\v{z}}ar et al. (\citeyear{dzudzar19a}) report that gas-rich galaxies that are part of groups also follow the $f_{atm} - q$ relation, indicating that the sAM of galaxies is an important driver of their \h1 gas even in group-like environments. Recently, Li et al. (\citeyear{Li20}) use a sample of galaxies from the VIVA survey (Chung et al.~\citeyear{Chung09}) to study the behaviour of a sub-sample of galaxies in the Virgo cluster on the $f_{atm} - q$ plane. They find that galaxies in their sample lie consistently below the relation, indicating that extreme environmental processes (such as ram pressure stripping) have removed significant fractions of their atomic gas without affecting their $q$ values.

Additionally, Romeo \& Mogotsi (\citeyear{romeo18}) discuss the role of AM and mass in regulating the local instabilities in galaxies, while a more generic stability parameter that connects the sAM of individual disk components - such as cold (H$_{2}$) and warm neutral medium (\h1) as well as stars - to their individual fractions (H$_2$, \h1 and stellar mass fractions) was introduced by Romeo (\citeyear{romeo20}). All these studies show the importance of AM in influencing star formation in disc galaxies, and how tightly it is linked to their global properties.

In this work, we extend previous studies that have explored the $f_{atm} - q$ parameter-space and populate it with the largest sample of galaxies to date, for which high-resolution \h1 data has been procured from the Westerbork \h1 Survey of Spiral and Irregular galaxies (WHISP; Swaters et al.~\citeyear{whisp02}). Robust rotation curves have been derived from 3D kinematic fitting to the \h1 data, which are then used to calculate precise values of their total baryonic AM. We examine the effects of the environment on the sample galaxies, probe their star formation properties, as well as study their morphology to establish a holistic understanding of the processes affecting their behaviour on the $f_{atm} -q$ plane.
The results from this study are highly relevant for the upcoming WALLABY \h1 survey (Koribalski et al.~\citeyear{Koribalski20}) using ASKAP, which has the potential to obtain high-resolution \h1 data for thousands of galaxies. The WALLABY survey will have similar spatial and spectral resolutions as that of the WHISP sample. This study will therefore enable us to make predictions and formulate expectations from the WALLABY survey and other future large \h1 surveys such as the SKA, in the context of using \h1 as a tool to probe the AM in galaxies.

In Section~\ref{Sec: Sample and methods}, we present the sample and discuss the methods employed in our AM analysis, as well as the different techniques used to probe the local environment. We present the main results and the following discussions in Section~\ref{Sec:Results}. Finally, we summarise the main results in Section~\ref{Sec:conclusions}. We have assumed the following cosmology for the current study: $\Omega = 0.27$, $\Lambda = 0.73$ and $H_{0} = 73$ \kms Mpc$^{-1}$.

\vspace{-0.4cm}
\section{Sample and methods}
\label{Sec: Sample and methods}

\subsection{The sample}
\label{Subsec:Sample}

\begin{table*}
    \centering
    \caption{The range of stellar and \h1 mass of the xGASS, WHISP and THINGS sub-samples for comparison. Also included are the mean and median of the masses.}
    \label{tab:xGASS-WHISP-stats}
    \begin{tabular}{llcccccc}
    \hline \hline
    Sample & Sample size & $M_{\star}$ range & Mean $M_{\star}$  & Median $M_{\star}$  & $M_{\textrm{\h1}}$ range & Mean $M_{\textrm{\h1}}$ & Median $M_{\textrm{\h1}}$ \\
        &   & [M$_{\odot}$] & [M$_{\odot}$] & [M$_{\odot}$]   & [M$_{\odot}$] & [M$_{\odot}$] & [M$_{\odot}$]    \\
        \hline
        WHISP & 114      & 6.7 -- 11.5 & 9.6 & 9.7 & 7.8 -- 10.5 & 9.3 & 9.3   \\
        \hline
        xGASS (detections & 803 & 9.0 -- 11.4 & 10.2 & 10.2 & 7.9 -- 10.5 & 9.4 & 9.5  \\
        only)       &      &      &      \\
        \hline
        THINGS & 16      & 9.1 -- 10.9 & 10.3 & 10.4 & 8.3 -- 10.1 & 9.5 & 9.6 \\
        \hline
        \end{tabular}
\end{table*}

The WHISP survey is one of the largest resolved \h1 surveys to date, with high-resolution data available for over 400 galaxies. The galaxies in the original WHISP sample were selected from the Uppsala General Catalogue of Galaxies (UGC; Nilson~\citeyear{nilson73}), and observed with the Westerbork Synthesis Radio Telescope (WSRT) with a synthesised beam $ \sim 14 \arcsec \times 14 \arcsec /sin\delta$. The target galaxies were required to have $D_{25} > 1.5 \arcmin$ (where $D_{25}$ is the B-band isophotal diameter at 25 mag arcsec$^{-2}$), with declinations north of 20\deg and additionally have \h1 peak flux densities $> 100$mJy. 

For this work, we make use of the $30\arcsec$ resolution data cubes (see Swaters et al.~\citeyear{whisp02}), with typical velocity resolutions of $\sim 5$ \kms. From the primary sample, we selected galaxies with \h1 disc radius spanning at least five resolution elements. This ensures that there are enough resolution elements to accurately fit 3D tilted-ring models to the galaxies and extract their kinematic properties. In addition, we selected only those galaxies with inclination angles between 20\deg and 80\deg, to avoid highly face-on and edge-on systems. The final sample consists of 114 galaxies (listed in Appendix \ref{appendix:properties_table} ). 

\begin{figure}
\hspace*{-0.5cm}
\includegraphics[width=8.2cm,height=6.2cm]{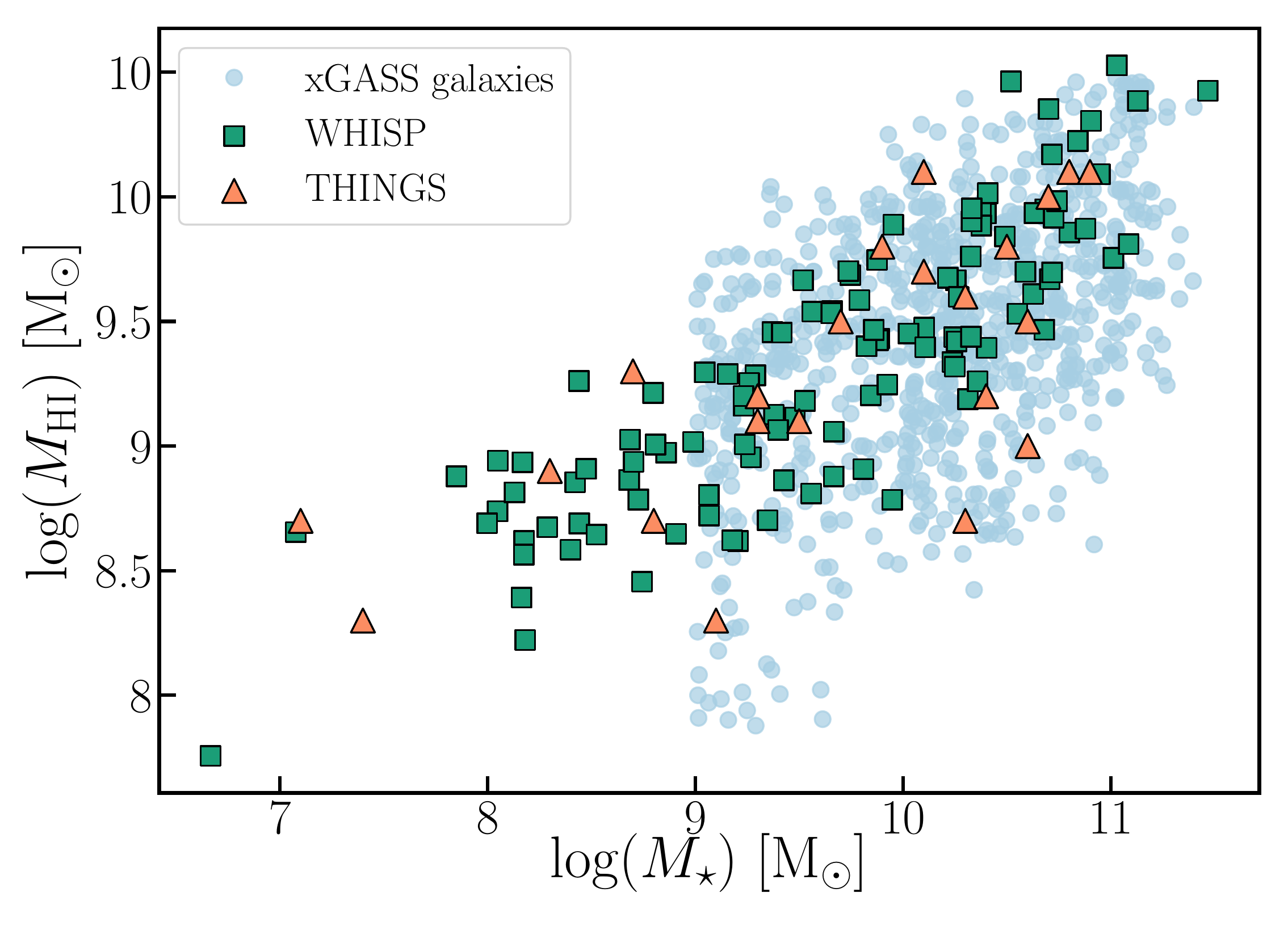}
\caption{The $M_{\textrm{\h1}} - M_{\star}$ scaling relation for the current WHISP sample (green squares) along with the xGASS representative sample in the background (light blue circles) for comparison. THINGS galaxies used in the original study by \citetalias{obreschkow14} are indicated by the orange triangles. The xGASS sample's stellar mass range is limited to $9 < \log M_{\star}/$M$_{\odot} < 11.5$. Within this range, our sample galaxies are observed to be uniformly spread in both stellar and \h1 mass.}
\label{fig1}
\end{figure}

\begin{figure}
\hspace*{-0.5cm}
\includegraphics[width=8.2cm,height=6.2cm]{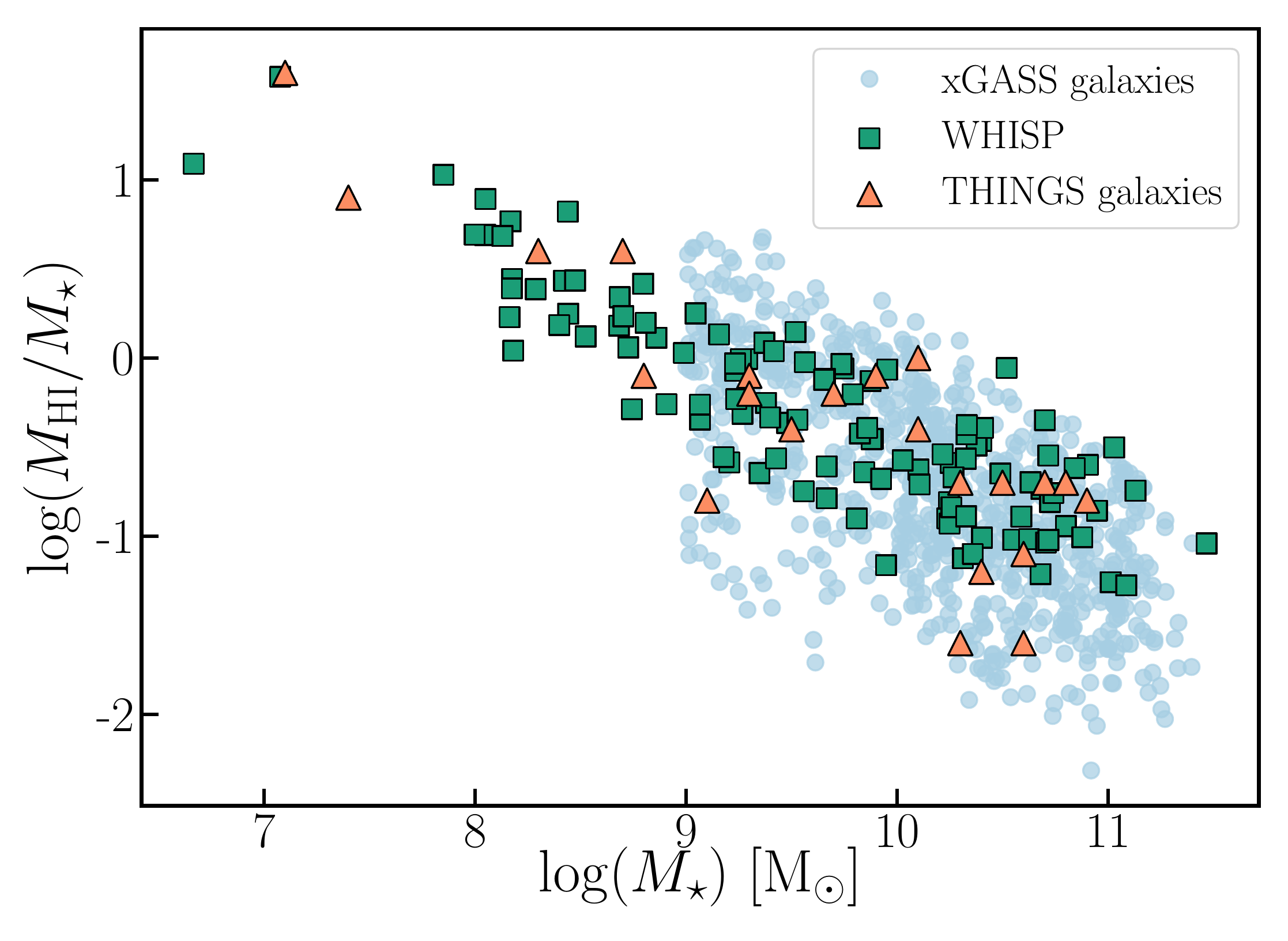}
\caption{The $M_{\textrm{\h1}}/M_{\star} - M_{\star}$ scaling relation for the WHISP sample (green squares). For comparison, also plotted is the xGASS sample (blue circles), while THINGS galaxies are represented by the orange triangles. We observe that our sample spans uniformly in both $M_{\star}$ and $M_{\textrm{\h1}}/M_{\star}$.}
\label{fig2}
\end{figure}

The stellar mass range for this sample varies between  $7 < \log M_{\star}/$M$_{\odot} < 11.5$. This is a wide range, spanning over five decades in stellar mass, including dwarfs, irregulars and spiral galaxies. We compare the \h1 and stellar mass properties of our sample with the xGASS representative sample (Catinella et al.~\citeyear{catinella18}). The xGASS survey is one of most sensitive single-dish extra-galactic \h1 surveys with \h1 detections for over 1000 nearby galaxies. It is an unbiased survey, in that, all galaxies within the stellar mass range $9 < \log M_{\star}/$M$_{\odot} < 11.5$ and within the redshift range $0.025 < z < 0.05$ are observed until \h1 is detected or a low gas-mass fraction (1.5 - 5\%) is reached. This makes the xGASS sample a gold standard for comparing the \h1 and stellar properties of nearby galaxies such as those in the WHISP sample. To see if our sample spans uniformly in both \h1 and stellar mass, we plot the $M_{\textrm{\h1}} - M_{\star}$ scaling relation alongside the xGASS galaxies for comparison, as shown in Fig.~\ref{fig1}. In addition to the xGASS galaxies, we also plot the 16 THINGS galaxies originally used by \citetalias{obreschkow14} and \citetalias{obreschkow16} in their study introducing the $f_{atm} - q$ relation. We see that our sample spans uniformly in both stellar and \h1 mass compared to the xGASS sample in the higher-mass end ($ \log M_{\star}/$M$_{\odot} > 9.0$). While we do not have a good reference sample in the lower-mass end ($ \log M_{\star}/$M$_{\odot} < 9.0$) to compare our sample galaxies with, based on the fact that the WHISP galaxies are \h1 selected, coupled with our size selection criteria means that the galaxies in the lower-mass end are likely to be biased towards gas-rich, low-mass spirals and/or dwarfs. Table.~\ref{tab:xGASS-WHISP-stats} lists some statistics for all three samples. Also shown in Fig.~\ref{fig2} is the \h1 gas fraction -- stellar mass scaling relation for the WHISP, xGASS and THINGS samples. We list all the relevant properties and derived quantities for our sample in Table.~\ref{tab:sample}

\vspace{-0.3cm}
\subsection{Total baryonic mass ($M_b$) and specific baryonic angular momentum ($j_b$)}

We perform 3D tilted-ring fitting (Rogstad et al.~\citeyear{rogstad74}) to the $30\arcsec$ resolution WHISP data cubes using \texttt{3DBarolo} (Di Teodoro et al.~\citeyear{teodoro15}) to model the galaxies and extract their kinematics. For every galaxy, a list of best initial guesses for the center, systemic velocity, inclination ($i$), position angle (PA), maximum rotation velocity and dispersion velocity ($\sigma_{\textrm{\h1}}$) are provided as input to the code. \texttt{3DBarolo} uses this information to generate model data cubes which are convolved with the synthesised beam of the instrument. Following this, a $\chi^{2}$ minimisation is performed ring-by-ring between the observed and model data cubes and a best fit model is determined. The output from the best fitting model encompasses a robust rotation curve, $\sigma_{\textrm{\h1}}$, $i$ and PA values for every ring, as well as the \h1 surface density profile of the galaxy. We make use of both the geometric and kinematic parameters resulting from the fit to then calculate the \h1 mass, stellar mass and total angular momentum within each ring.

Following the methods described in Murugeshan et al. (\citeyear{murugeshan2019}), we project the tilted-rings onto the moment-0 \h1 intensity maps to compute the \h1 mass within each ring. To calculate the stellar mass, we make use of the 2MASS (Skrutskie et al.~\citeyear{2mass06}) $K_{s}$-band mosaics, after carefully masking foreground stars and performing a background sky subtraction. The sum of the $K_s$ magnitudes within each ring is then converted to a stellar mass following the relation described by Eq.3 in Wen et al. (\citeyear{wen13}) as follows 
\begin{equation*}
    \begin{aligned}
\log _{10}\left(\frac{M_{\star}}{\mathrm{M}_{\odot}}\right)=&(-0.498 \pm 0.002)+(1.105 \pm 0.001) \\
& \times \log _{10}\left(\frac{v L_{v}\left(K_{s}\right)}{\mathrm{L}_{\odot}}\right)
\end{aligned}
\end{equation*}
Where $L_{v}$ is the luminosity, derived using the extinction-corrected $K_s$-band magnitude.

The total \h1 and stellar masses are computed by summing their respective mass within each ring. We then compute the total baryonic mass using the relation $M_b = M_{\star} + 1.35(M_{\small \mathrm{\textrm{\h1}}} + M_{\small \mathrm{H_{2}}})$, where $M_{\small \textrm{\h1}}$ is the total \h1 mass, $M_{\star}$ is the total stellar mass and $M_{\small \mathrm{H_{2}}}$ is the H$_{2}$ mass. The factor 1.35 accounts for the universal 26\% He fraction. Due to a lack of CO data for the sample, we have assumed $M_{\small \mathrm{H_{2}}}/M_b \sim 4\%$ following the observations made by Obreschkow \& Rawlings (\citeyear{obreschkow09}) for a number of local late-type galaxies.
The total specific baryonic angular momentum is computed as 
\begin{equation}
    j_b =\frac{\sum_{i} (1.35M_{\mathrm{\tiny \textrm{\h1}},i}+M_{\star,i})V_{rot,i}r_{i}}{\sum_{i} (1.35M_{\mathrm{\tiny \textrm{\h1}},i}+M_{\star,i})}
\end{equation}
where $r_{i}$ is the radius of the $i^{th}$ ring and $V_{rot,i}$ is the rotation velocity corresponding to that ring. Following this, as mentioned in Section~\ref{Sec:Intro}, we compute the atomic gas fraction and stability parameter as
\begin{equation*}
    f_{atm} = \frac{1.35 M_{\textrm{\h1}}}{M_b}; \hspace{0.5cm} q = \frac{j_{b}\sigma}{G M_b}
\end{equation*}

\subsection{Probing the environment}
\label{Subsec:Environment}

In order to understand the influence of the environment on our sample galaxies, we make use of methods that probe and quantify the effects of both the very local (interacting pairs; close companions) and intermediate environments (group to inter-group regime) of the sample galaxies. Our sample consists of isolated galaxies and those that are in pairs, triplets and some that are part of groups. As such, environmental factors are likely to be playing an important role in re-distributing the \h1 gas in such systems. To study the influence of close neighbours/companions on the sample galaxies, we divided the sample into two further sub-samples -- galaxies with close neighbours and those without. A galaxy is considered to have a close neighbour if one or more companion galaxies are found within a projected distance of 200 kpc and $\pm ~250$ \kms ~in systemic velocity of the galaxy. We make use of NED's\footnote{NASA/IPAC Extragalactic Database, \url{http://ned.ipac.caltech.edu/}} environment search, and also examine the \h1 image cubes to identify close neighbours. 

We probe the density of the intermediate environment of the sample galaxies using the projected nearest-neighbour density metric ($\Sigma_{\textrm{N}}$ [Mpc$^{-2}$]). The $\Sigma_{\textrm{N}}$ values have been computed using the 2MASS Redshift Survey (2MRS) catalogue (Huchra et al.~\citeyear{huchra12}). The 2MRS catalogue contains measured spectroscopic redshifts for over 43,500 galaxies with $K_s \leq 11.75$ mag and |b| > 5\deg. Within these limits, the survey is complete to 97.6\% and covers 91\% of the entire sky. The 2MRS is, however, a relatively shallow survey and the magnitude limits are based on the $K_s$-band magnitudes. Therefore, the catalogue by virtue of the selection criteria will be more sensitive to galaxies that are older and redder. In-order to make the 2MRS catalogue volume-limited for an unbiased measurement of the local densities, we employ the following two steps:
\begin{enumerate}
    \item We first make a velocity cut to the original 2MRS catalogue, by selecting galaxies within the velocity range 200 -- 8000 \kms. 
    \item To this velocity-cut sample, we add an absolute $K_{s}$-band magnitude cut $M_K < -23.45$, corresponding to the survey's limiting apparent magnitude of 11.75 mag at the highest velocity/redshift edge (8000 \kms). This makes the 2MRS sample volume-limited.
\end{enumerate}

Finally, we exclude galaxies in our WHISP sample which have systemic velocities $V_{sys} < 700$ \kms ~to avoid peculiar velocity effects.
This reduces our WHISP sample size from 114 to 91. Since we have imposed the above cuts to the original 2MRS catalogue to make it volume-limited, the final reference catalogue will not include many low-surface brightness galaxies within the volume, making the distribution of galaxies in the final reference catalogue sparse. For this reason, following the justifications made by Janowiecki et al. (\citeyear{janowiecki19}), we use the second nearest-neighbour density metric ($\Sigma_2$) so that we are sensitive to only the intermediate environment as otherwise the $\Sigma_3$ and/or $\Sigma_5$ metrics are likely to be probing large-scale structures. The $\Sigma_2$ metric is defined as follows $\Sigma_2 = 2/\pi D^2$, where $D$ is the projected distance to the 2nd nearest-neighbour within $\pm ~500$\kms. In addition to the local environment densities, we use the 2MRS group catalogue (Lu et al.~\citeyear{Lu16}) to study the effects of group membership on the \h1 gas fraction and angular momentum properties of the sample galaxies. 

\vspace{-0.4cm}
\section{Results and Discussion}
\label{Sec:Results}

We now present the results from the study. This work presents the largest sample of late-type galaxies for which accurate baryonic angular momentum ($j_b$) has been computed to date. In the following subsections we present the distribution of $j_b$ for our sample galaxies and determine if the $j_b$ values vary with their environment. Additionally, we examine the behaviour of galaxies in the $f_{atm} - q$ plane and study the influence of the environment on their position in this parameter-space. We also discuss the trends of other global properties of the sample galaxies such as their star formation rate (SFR), star formation efficiency (SFE) and bulge-to-total (B/T) ratios in the $f_{atm} - q$ parameter-space.

\vspace{-0.3cm}
\subsection{The $j_b - M_b$ relation}
\label{Sec:Result j_b distribution}

\begin{figure}
\hspace*{-0.5cm}
\includegraphics[width=8.3cm,height=6.3cm]{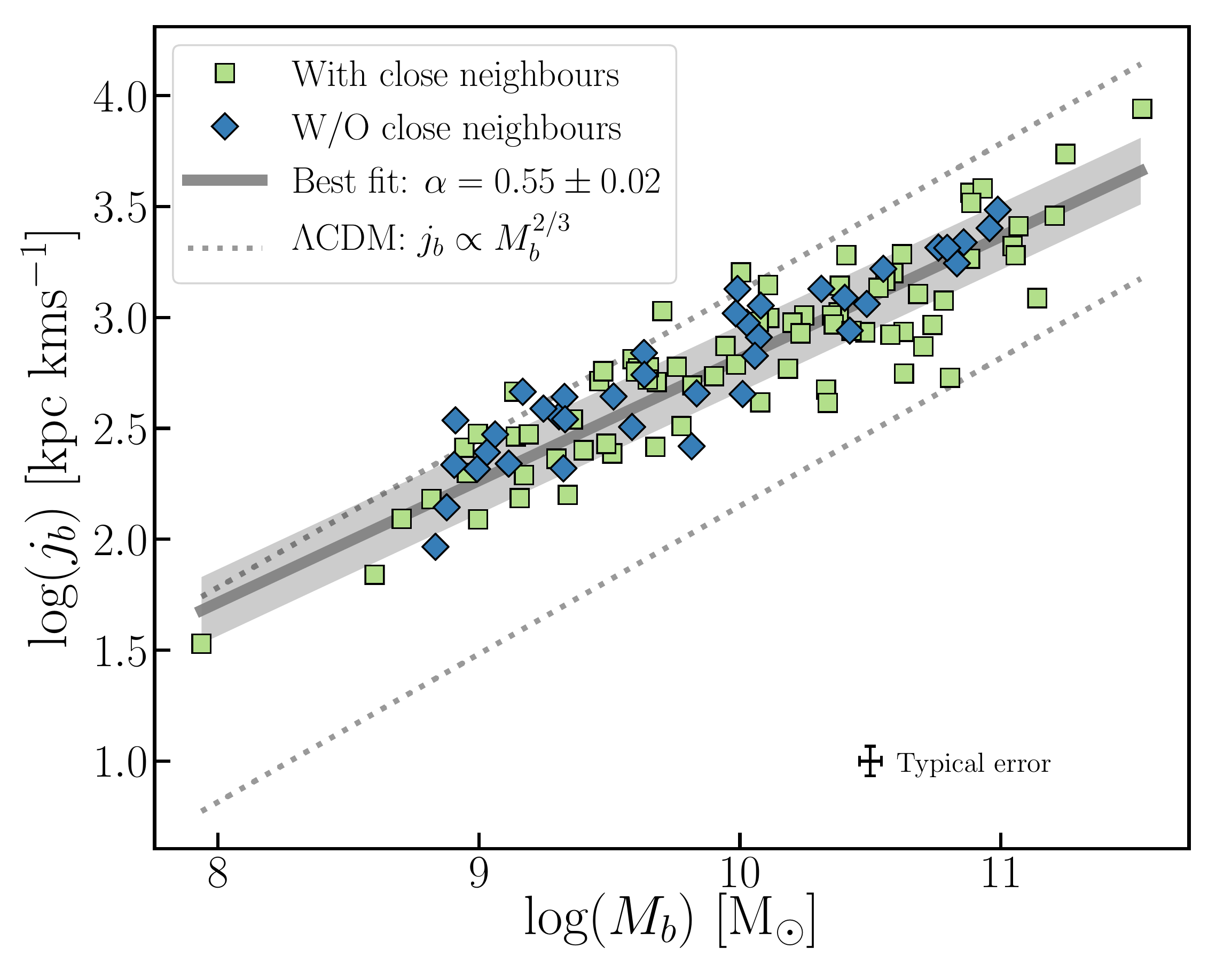}
\caption{Distribution of the total specific baryonic angular momentum ($j_b$) as a function of total baryonic mass ($M_b$) for the 114 WHISP galaxies. The dotted lines show the expected range of $j_b$ values based on $\Lambda$CDM predictions. The solid dark gray line is the best fitting line to the entire sample with a slope $\alpha = 0.55 \pm 0.02$. The light grey region is the $1 \sigma$ intrinsic scatter of $\sim 0.13 \pm 0.01$ dex for the full sample.}
\label{fig3}
\end{figure}

In this section we discuss the distribution of $j_b$ as a function of $M_b$ for our sample of 114 WHISP galaxies. As described in Section~\ref{Subsec:Environment}, the sample was further divided into two -- those with close neighbours and those without. Fig.~\ref{fig3} shows the $j_b - M_b$ relation for the full sample. We fit a linear regression\footnote{Fitting was performed using Hyper-Fit, an R package for fitting multi-dimensional data. See Robotham \& Obreschkow~\citeyear{Robotham15} for more details.} of the form
\begin{equation*}
    \log j_b = \alpha \log M_b + c
\end{equation*}
to the full sample as well as separately for the two sub-samples. For the full sample, we find a best fitting slope of $\alpha = 0.55 \pm 0.02$ (dark grey line in Fig.~\ref{fig3}) and a $1\sigma$ intrinsic scatter $\sim 0.13 \pm 0.01$ dex (light grey shaded region in the plot). For the sample of galaxies without close neighbours we find a slope of $\alpha = 0.54 \pm 0.04$ and an intrinsic scatter $\sim 0.10 \pm 0.02$ dex. The sample of galaxies with close neighbours have a slope $\alpha = 0.56 \pm 0.03$, but show a slightly larger intrinsic scatter of $\sim 0.14 \pm 0.02$ dex. It is interesting to note that the slopes of the two sub-samples are statistically consistent with each other, however, the larger observed scatter for the sample with close neighbours is possibly a result of the effects of galaxy interactions on their $j_b$ values.

It is also worth noting that some galaxies with close neighbours (indicated by the green squares in Fig.~\ref{fig3}), particularly in the higher baryonic mass end, have significantly low $j_b$ compared to galaxies without close neighbours of similar baryonic mass, indicating that past or ongoing interactions are likely to have lowered their total specific baryonic AM. Many of these outliers in the $j_b - M_b$ relation, located beyond the $1 \sigma$ scatter (light grey region), are also outliers in the $f_{atm} - q$ relation (see Section~\ref{Sec:fatm-q-effects_of_close_neighbours}), indicating a strong connection between specific AM and disc stability. We discuss this further in Section~\ref{Sec:fatm-q-effects_of_close_neighbours}.

\citetalias{obreschkow14} derive a theoretical relation between $j_b$ and $M_b$ (based on the original prescriptions of Mo et al.~\citeyear{Mo98}) for the local universe (assuming $H = 70$~\kms Mpc$^{-1}$) of the form
\begin{equation}
    \frac{j_{\mathrm{b}}}{10^{3} \mathrm{kpc} ~\mathrm{km} \mathrm{s}^{-1}}=1.96 \lambda f_{\mathrm{j}} f_{\mathrm{M}}^{-2 / 3}\left(\frac{M_{\mathrm{b}}}{10^{10} \mathrm{M}_{\odot}}\right)^{2 / 3}
\end{equation}
Where $\lambda$ is the dimensionless halo spin parameter, $f_{\mathrm{j}} = j_b/j_h$ is the fraction of sAM retained by the baryons and is the ratio of the specific baryonic angular momentum ($j_b$) and the specific angular momentum of the halo $j_h$. $f_{\mathrm{M}}$ is the baryon mass fraction. \citetalias{obreschkow14} assume $\lambda \approx 0.04 \pm 0.02$, $f_{\mathrm{j}} \approx 1$ (within about 50\%) and $f_{\mathrm{M}} \approx 0.05$ for late-type galaxies and show that the pre-factor $1.96 \lambda f_{\mathrm{j}} f_{\mathrm{M}}^{-2 / 3}$ will vary between 0.14 and 1.3 in the local universe (for more details see Section 4 in \citetalias{obreschkow14}). This range is shown by the dotted lines in Fig.~\ref{fig3}. We find that most galaxies in our sample lie within the expected range of $j_b$ values for their mass, but the slope of the relation is significantly lower than the expected slope of $\sim 2/3$ from the models of \citetalias{obreschkow14}. This can be explained by arguing that $f_{\mathrm{j}} f_{\mathrm{M}}^{-2 / 3}$ is not a constant for galaxies of all types and masses. In fact, Chowdhury \& Chengalur (\citeyear{Chowdhury17}) examine five gas-rich dwarfs and find that their $j_b$ is elevated compared to the model predictions for high-mass spirals. They suggest that this is primarily due to the mass dependence of $f_{\mathrm{M}}$, which decreases with decreasing baryonic mass (see for example Crain et al.~\citeyear{Crain07}). 

In an independent study, Butler, Obreschkow \& Oh (\citeyear{butler17}) come to a similar conclusion after analysing the $j_b - M_b$ relation for 14 dwarf galaxies from the LITTLE THINGS survey. In addition, both Chowdhury \& Chengalur (\citeyear{Chowdhury17}) and Kurapati et al. (\citeyear{Kurapati18}) suggest that $f_{\mathrm{j}}$ is also likely to vary with mass based on preferential `cold-mode' accretion among low-mass systems compared to high-mass galaxies that tend to accrete in `hot-mode'. Many simulations allude to the fact that cold-mode accretion is associated with high AM gas as opposed to hot-mode accretion, thereby boosting the sAM of low-mass spirals (Pichon et al.~\citeyear{Pichon11}; Stewart et al.~\citeyear{Stewart11}; Danovich et al.~\citeyear{danovich15}). 

Furthermore, Posti et al. (\citeyear{Posti18}) made an empirical study of the $j_{\star} - M_{\star}$ relation (the Fall relation) for late-type galaxies, that included dwarfs and high-mass spirals spanning over 5 decades in stellar mass. In this study, they find that their sample galaxies follow an unbroken single power-law of the form $j_{\star} \propto M_{\star}^{0.55 \pm 0.02}$, and point to the fact that this can be explained by prescribing a biased collapse model (see for example Kassin et al.~\citeyear{Kassin12}; Dutton \& van den Bosch~\citeyear{dutton12}), where $f_j$ decreases with decreasing mass. In addition, using a suite of cosmological zoom-in simulations from the FIRE project, El-Badry et al. (\citeyear{El-Badry18}) show that by introducing strong stellar feedback in low-mass systems, $f_j$ can be reduced significantly in low-mass galaxies.

Interestingly, the slope we derive for the $j_b - M_b$ relation from our full sample ($\alpha = 0.55 \pm 0.02$) matches exactly with the best fitting slope of Posti et al. (\citeyear{Posti18}), thus strengthening the argument in favour of variation in both $f_{\mathrm{M}}$ and $f_{\mathrm{j}}$. Romanowsky \& Fall (\citeyear{romanowsky12}) in their study of the $j_{\star} - M_{\star}$ plane find a slope $\alpha \approx 0.52$ for their sample when including both disc and bulge components, again consistent with our findings. 

In another study of the $j_b - M_b$ relation, Elson (\citeyear{Elson17}) use 37 galaxies from the WHISP sample and find a best fitting slope of $\alpha = 0.62 \pm 0.02$, in disagreement with the best fitting slope for our full sample. We suspect the apparent discrepancy with the Elson study may be due to two main reasons. Firstly, the mass distribution of the two samples are different. Elson specifically probe low-mass spirals ($8 < \log M_b < 10$ [M$_{\odot}$]) in their study, while the baryonic mass range for our sample is $8 < \log M_b < 11.5$ [M$_{\odot}$] with a significant fraction (50\%) of galaxies in the high-mass ($\log M_b > 10$ [M$_{\odot}$]) end. To rule-out any biases introduced from the different methodologies employed in the measurement of $j_b$ and $M_b$ values in the two studies, we also compared the $j_b$ and $M_b$ values for an overlapping sample of 26 WHISP galaxies and find a good one-to-one correlation. We then fit a line to the $j_b - M_b$ relation for these 26 common galaxies, and find a best fitting line with slope $\alpha = 0.63 \pm 0.05$, consistent with the results from the Elson study. We therefore believe, that the mass ranges probed in the two studies play an important role in affecting the fitted slopes. Secondly, because our sample is size-selected (see Section~\ref{Subsec:Sample}), we may be naturally biased towards large high-spin, gas-rich systems, especially in the dwarf regime. As a consequence, the slope of the $j_b - M_b$ relation ($\alpha$) becomes flatter due to this selection bias. A combination of these reasons may explain the observed discrepancy between the best fitting slopes of the two samples. 

The flattened and unbroken power-law we observe for our sample has important implications for current galaxy evolution models, particularly how different types of galaxies acquire, retain or lose their AM, and the effects of feedback processes on the fraction of baryons retained during galaxy formation. However, better statistical constraints of this relation would require a larger and more homogeneous sample than that considered in this work, designed to study environmental effects, and not prone to any selection biases (e.g. size, mass, flux selected). 

\vspace{-0.3cm}
\subsection{Effects of close neighbours}
\label{Sec:fatm-q-effects_of_close_neighbours}

\begin{figure}
    \hspace*{-0.5cm}
    \includegraphics[width=8cm,height=6.5cm]{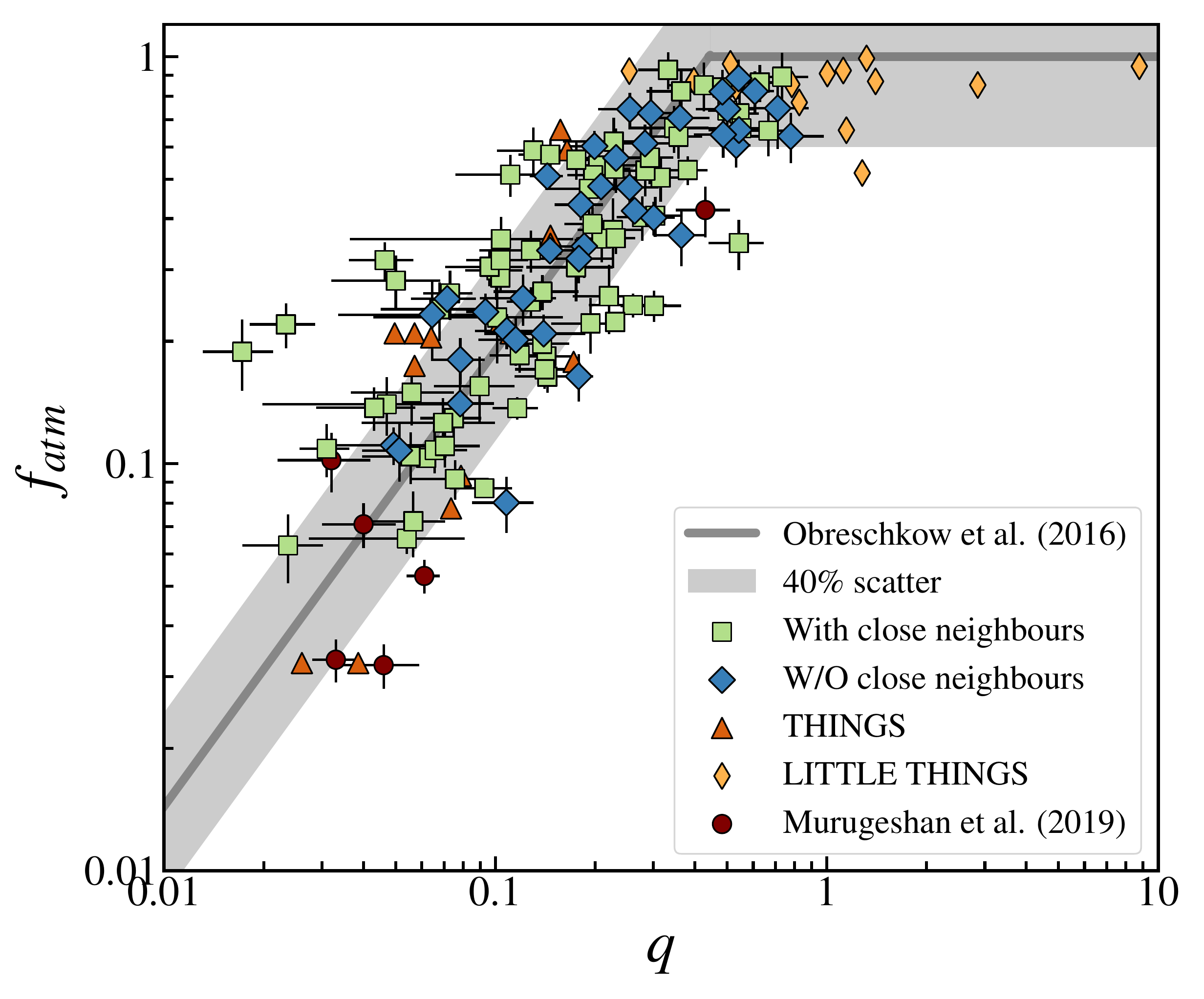}
    \caption{The $f_{atm} - q $ relation for 114 WHISP galaxies. Blue diamonds are WHISP galaxies with no close neighbours and light green squares are galaxies in the WHISP sample with close neighbours (see Section~\ref{Subsec:Environment}). Galaxies in the sample without close neighbours are seen to follow the model very consistently (scatter $\sim 0.13$ dex), whereas galaxies with close neighbours are seen to have a larger scatter ($\sim 0.22$ dex) about the relation with clear outliers. The light grey region is the 40\% empirical intrinsic scatter in the \h1 dispersion velocities ($\sigma_{\textrm{\h1}}$) of galaxies. Also plotted for reference is the original sample of THINGS and LITTLE THINGS galaxies from \citetalias{obreschkow16} and the \h1-deficient spiral galaxies from Murugeshan et al. (\citeyear{murugeshan2019})}
    \label{fig4}
\end{figure}

We now take a look at the effects of the local environment on the \h1 gas, specific angular momentum and the stability parameter $q$ of our sample galaxies, and how this affects their behaviour on the $f_{atm} - q$ relation. Fig.~\ref{fig4} shows the relation colour-coded into galaxies that have close neighbours (green squares) and those that do not (blue diamonds). Galaxies without close neighbours have an rms scatter of $\sim 0.13$ dex about the relation (note that this is smaller than the rms scatter in the original work by \citetalias{obreschkow16}, which was 0.2 dex), while galaxies with close neighbours show an rms scatter of $\sim 0.22$ dex, almost a factor of two compared to the sample without close neighbours. This larger scatter may be linked to galaxy-galaxy interactions and its effects on their $j_b$ and $f_{atm}$ values. 

Interactions between galaxies tend to exert additional tidal torques on their gas discs, which lowers their internal specific AM by spin-orbit interaction, leading to the funnelling of gas to their centres. The cumulative effect of this is a net reduction of $j_b$ in both interacting galaxies (Barnes \& Hernquist~\citeyear{barnes96}), which lowers their $q$ value. Additionally, more massive galaxies will tend to accrete \h1 gas from their less massive gas-rich companions, momentarily boosting their gas fractions, $f_{atm}$ (see for example Ellison et al.~\citeyear{ellison18}). This will make the galaxy appear more \h1 rich for its given stability parameter ($q$). Both these effects collectively move interacting galaxies to the left and/or above the $f_{atm} - q$ relation. It has been shown that interacting pairs and galaxies with close companions show enhanced star formation, due to disc instabilities set by the external tidal perturbation (Ellison el al.~\citeyear{ellison08}; Ellison et al.~\citeyear{ellison10}; Patton et al.~\citeyear{patton11}; Scudder et al.~\citeyear{Scudder12}). To test this, in Section~\ref{Sec:fatm-q-SFR/SFE} we also examine if the SFR is elevated for those galaxies having close companions and deviating from the relation (see Fig.~\ref{fig9}).

This is an important result of this study, and motivates us to explore the use of the $f_{atm} - q$ relation as a diagnostic plot to identify galaxies having undergone or currently undergoing tidal interactions. This will become particularly important in high-density environments such as clusters, where the two main gas stripping mechanisms -- tidal and ram pressure stripping can be distinguished based on the location of galaxies on the $f_{atm} - q$ plane. Tidal interactions are likely to move galaxies to the left of the relation, while a fast gas stripping process like ram pressure, will move the galaxies below the relation (Li et al.~\citeyear{Li20}). Upcoming large \h1 surveys such as WALLABY (Koribalski et al.~\citeyear{Koribalski20}), will enable us to populate thousands of galaxies on this diagnostic plot and allow us to study the effects of both tidal interactions and ram pressure stripping on their $f_{atm}$, $j_b$ and $q$ values. 
 
\vspace{-0.3cm}
\subsection{Effects of the intermediate environment}
\label{Sec:fatm-q-environment}

In this section we examine the effects of the intermediate environment on the \h1 gas fractions, angular momentum and $q$ values of the sample galaxies. We make use of the methods described in Section~\ref{Subsec:Environment} to probe the environment density of the sample galaxies.

\begin{figure}
    \hspace*{-0.5cm}
    \includegraphics[width=8cm,height=6.2cm]{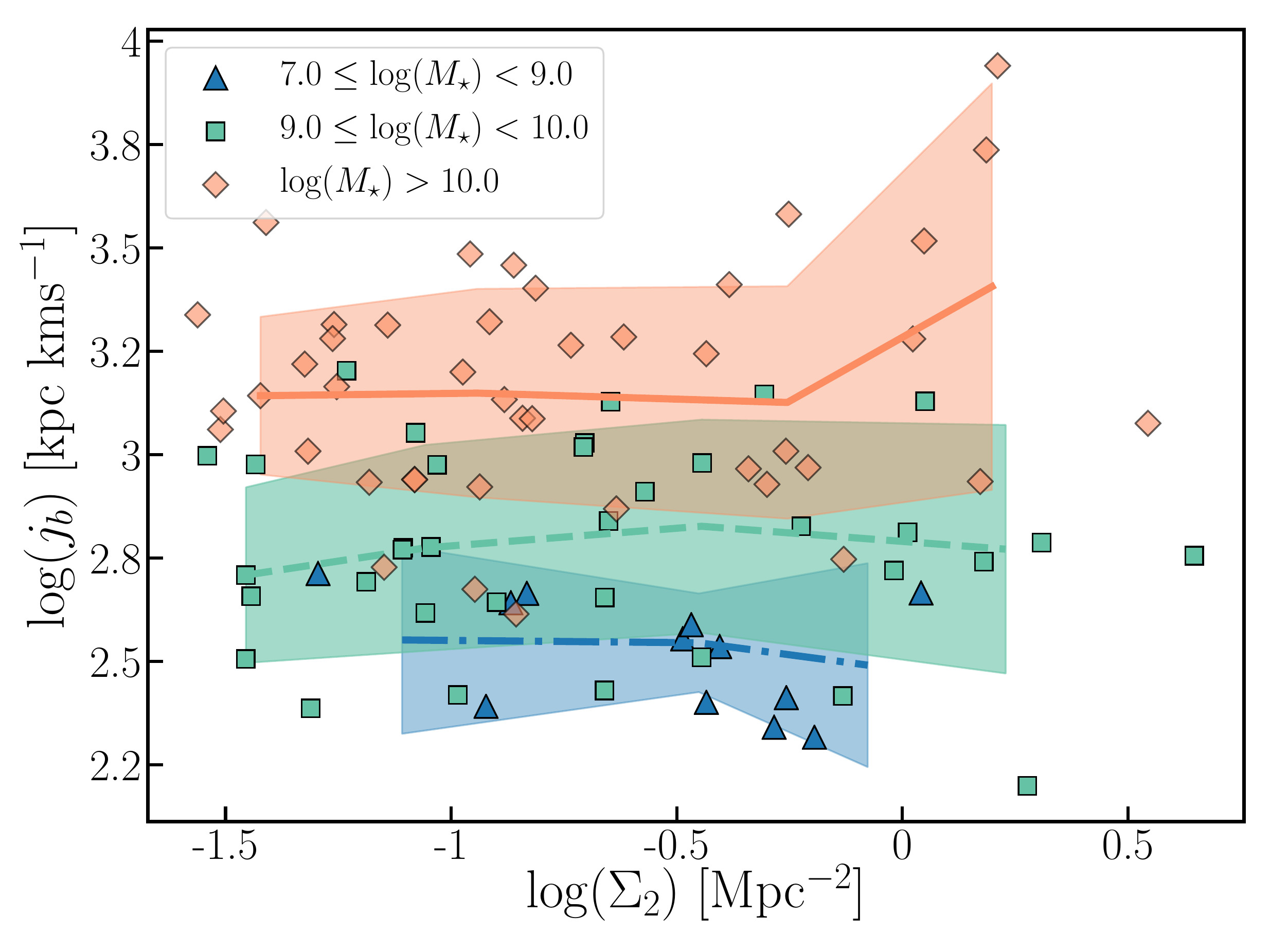}
    \caption{The total specific baryonic AM ($j_b$) is plotted against the 2nd nearest-neighbour density $\Sigma_{2}$ (for a sub-sample 91 galaxies) for three stellar mass bins -- $7.0 \leq \log(M_{\star}) < 9.0$ (blue triangles), $9.0 \leq \log(M_{\star}) < 10.0$ (green squares) and $\log(M_{\star}) > 10.0$ (orange diamonds). The different lines show the rolling medians for the different mass bins (blue dash-dot, green dash-dash and solid orange respectively). The shaded regions show the $1\sigma$ scatter about the rolling median. No trend between $j_b$ and $\Sigma_2$ is observed for the galaxies in all three stellar mass bins.}
    \label{fig5}
\end{figure}

We plot the $j_b$ values for our sub-sample of 91 galaxies against their $\Sigma_2$ density metric for three stellar mass bins as shown in Fig.~\ref{fig5}. We find no correlation between $j_b$ and $\Sigma_2$, indicating that the sAM of our current sample is not affected significantly by their intermediate environment. However, caution is warranted as this result by no means implies that $j_b$ does not vary with environment in general. This may simply be a selection effect, as our sample galaxies are drawn from the parent WHISP sample, which is heterogeneous and does not include galaxies from cluster-like environments, where interactions are more common and more low sAM systems are observed.

\begin{figure}
    \hspace*{-0.5cm}
    \includegraphics[width=8.5cm,height=6.5cm]{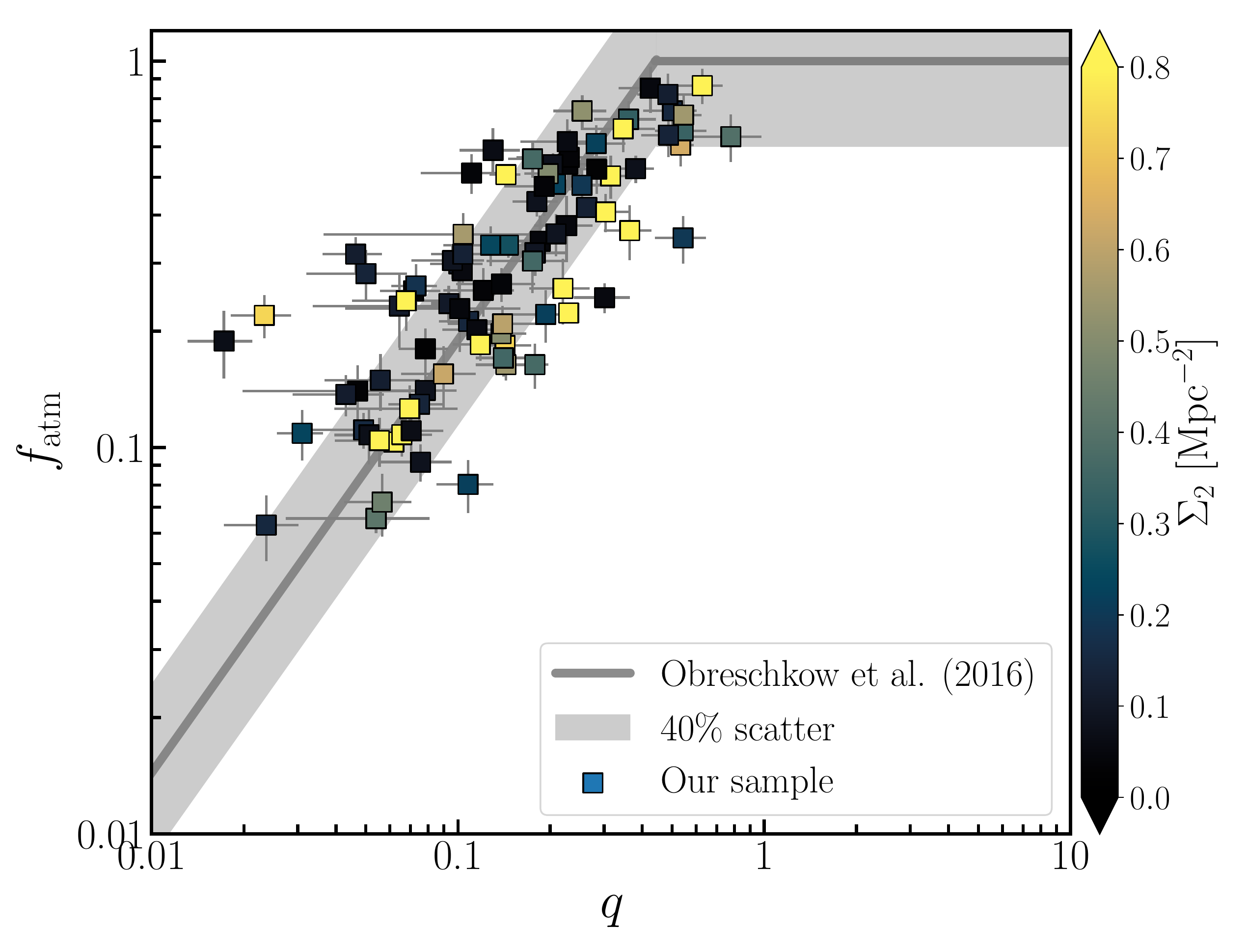}
    \caption{The $f_{atm} - q $ relation for a sub-sample of 91 galaxies, color coded by their environment densities, denoted by $\Sigma_2$. Overall, no trend is observed in the parameter-space indicating that the intermediate environment of the sample galaxies does not have a significant impact on their $f_{atm}$ and $q$. Most galaxies are from low- to intermediate-density environments. It is worth noting that all galaxies that are clear outliers (to the left of the $f_{atm} - q$ relation) are from low-density environments. These galaxies are in fact isolated interacting pairs and triplets.}
    \label{fig6}
\end{figure}

Fig.~\ref{fig6} shows the $f_{atm} - q $ relation for the sample galaxies, color coded by their $\Sigma_2$ values. Interestingly, the intermediate environments do not seem to have an effect on the stability of galaxies and their gas fractions. This indicates that in the group regime, unless close flybys and/or gas stripping is evident, galaxies follow the $f_{atm} - q$ relation consistently. It is also interesting to note that galaxies with close companions which are also outliers in the plot (deviating to the left of the relation), are all isolated pairs and triplets residing in relatively low-density environments. This again points to the fact that galaxies with close neighbours and undergoing gravitational interactions are most affected in terms of their disc stability and their position in the $f_{atm} - q$ plane. However, it is worth noting that in a recent work, Li et al. (\citeyear{Li20}) study the behaviour of cluster galaxies on the $f_{atm} - q$ plane. They make use of galaxies part of the VIVA survey (Chung et al.~\citeyear{Chung09}) and find that the galaxies are located below the relation, and are found to have significantly lower atomic gas fractions than expected for their $q$ values, consistent with what semi-analytic simulations predict (Stevens et al.~\citeyear{Stevens18}). This can be explained by the fast gas-stripping associated with ram pressure in clusters, which strips off the \h1 gas in the outskirts, but does not affect the total specific baryonic angular momentum of the galaxy. The net result is a drastic reduction in the atomic gas fraction while the $q$ value remains more or less preserved. We reiterate that such drastic reductions in \h1 gas fractions are not observed in our sample as they are mostly from isolated or group-like environments.

\vspace{-0.3cm}
\subsection{Group membership and halo mass effects}
\label{Sec:fatm-q-group_membership}

In this section we examine the behaviour of galaxies in our sample on the $f_{atm} - q$ plane in the context of their group membership. How do isolated centrals, galaxies in pairs/triplets and those part of groups behave in this parameter space? To understand this, we make use of the 2MRS group catalogue, published by Lu et al. (\citeyear{Lu16}) to identify isolated centrals, pairs, triplets and group galaxies in our sample. We first identify and cross-match galaxies in our sample and those in the group catalogue, each of which is assigned a unique galaxy ID and a group ID. Using this group ID, we then extract the group membership and associated halo mass details. This reduces our sample from 114 to 63. 

\begin{figure}
    \hspace*{-0.5cm}
    \includegraphics[width=8.5cm,height=6.5cm]{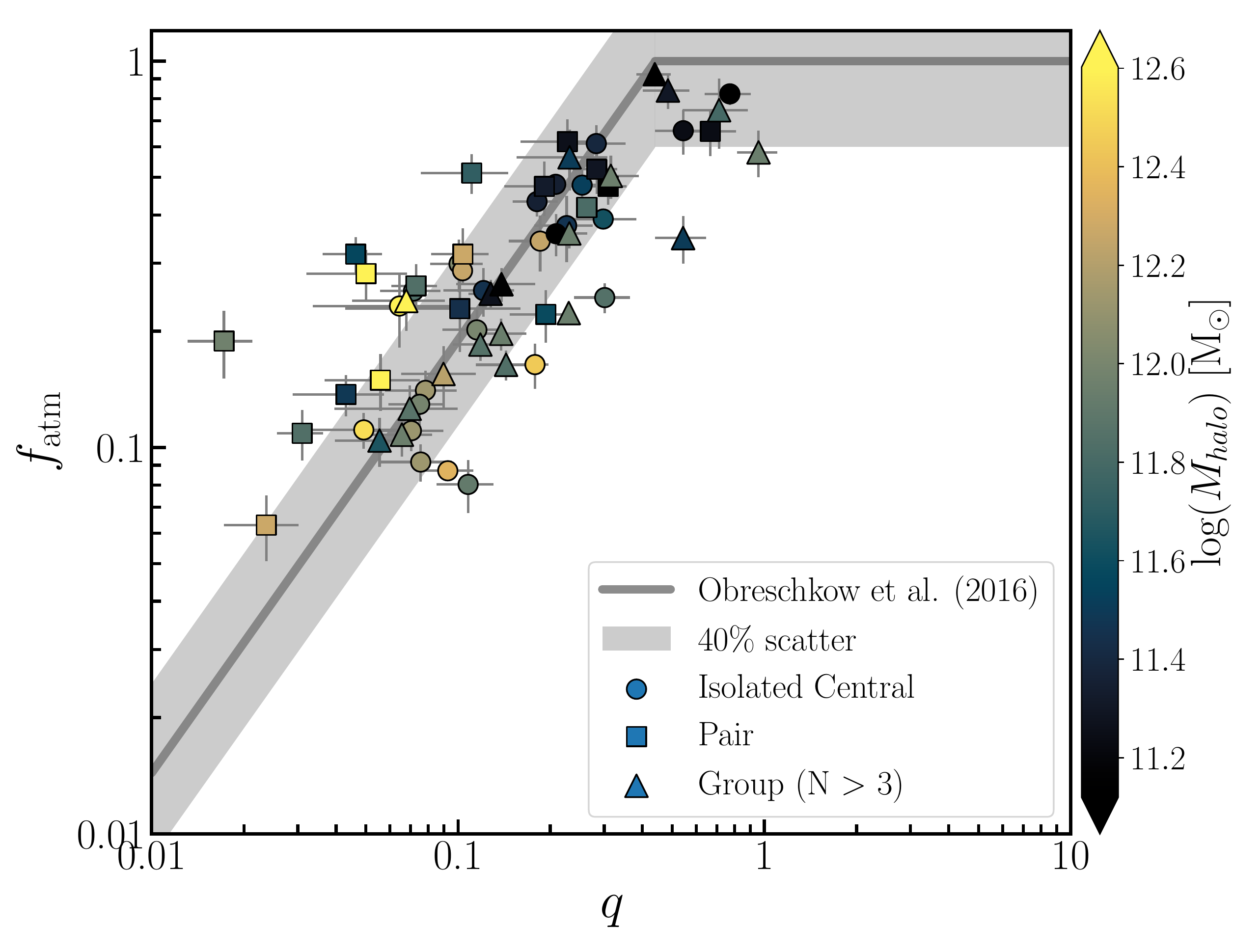}
    \caption{The plot shows the $f_{atm} - q$ relation for a sub-sample of 63 galaxies that are identified as isolated centrals, pairs and those belonging to triplets/groups, after cross-matching with the group catalogue by Lu et al. (\citeyear{Lu16}). Many galaxies identified as isolated centrals (circles) in the group catalogue are observed to be isolated pairs with small dwarf-like companions. In the plot, we observe that isolated centrals and pairs (squares) deviate the most from the relation, suggesting that interacting galaxies are the ones most affected and drift away from the relation.}
    \label{fig7}
\end{figure}

Fig.~\ref{fig7} shows a plot of the relation with the different symbols indicating the different environments the galaxies are part of, with isolated centrals, pairs and triplets and/or group galaxies represented by circles, squares and triangles respectively. The halo mass associated with the galaxies are shown in the color-bar. Interestingly, we find that galaxies identified as isolated centrals (circles) and as pairs (squares) are the ones deviating from the relation the most. Those galaxies identified to be part of groups are observed to follow the relation consistently, with a few exceptions. However, it is worth noting that many galaxies in the catalogue identified as isolated centrals by Lu et al. (\citeyear{Lu16}), do in fact have low surface brightness dwarf-like companions detected in \h1, missed possibly because the 2MRS is $K_s$-band selected. This observation again indicates that galaxies tend to drift away from the $f_{atm} - q$ relation due to the gravitational influence of their close neighbours. 

\vspace{-0.3cm}
\subsection{Relationship with stellar mass}
\label{Sec:fatm-q-M}
\begin{figure}
    \hspace*{-0.5cm}
    \includegraphics[width=8.5cm,height=6.5cm]{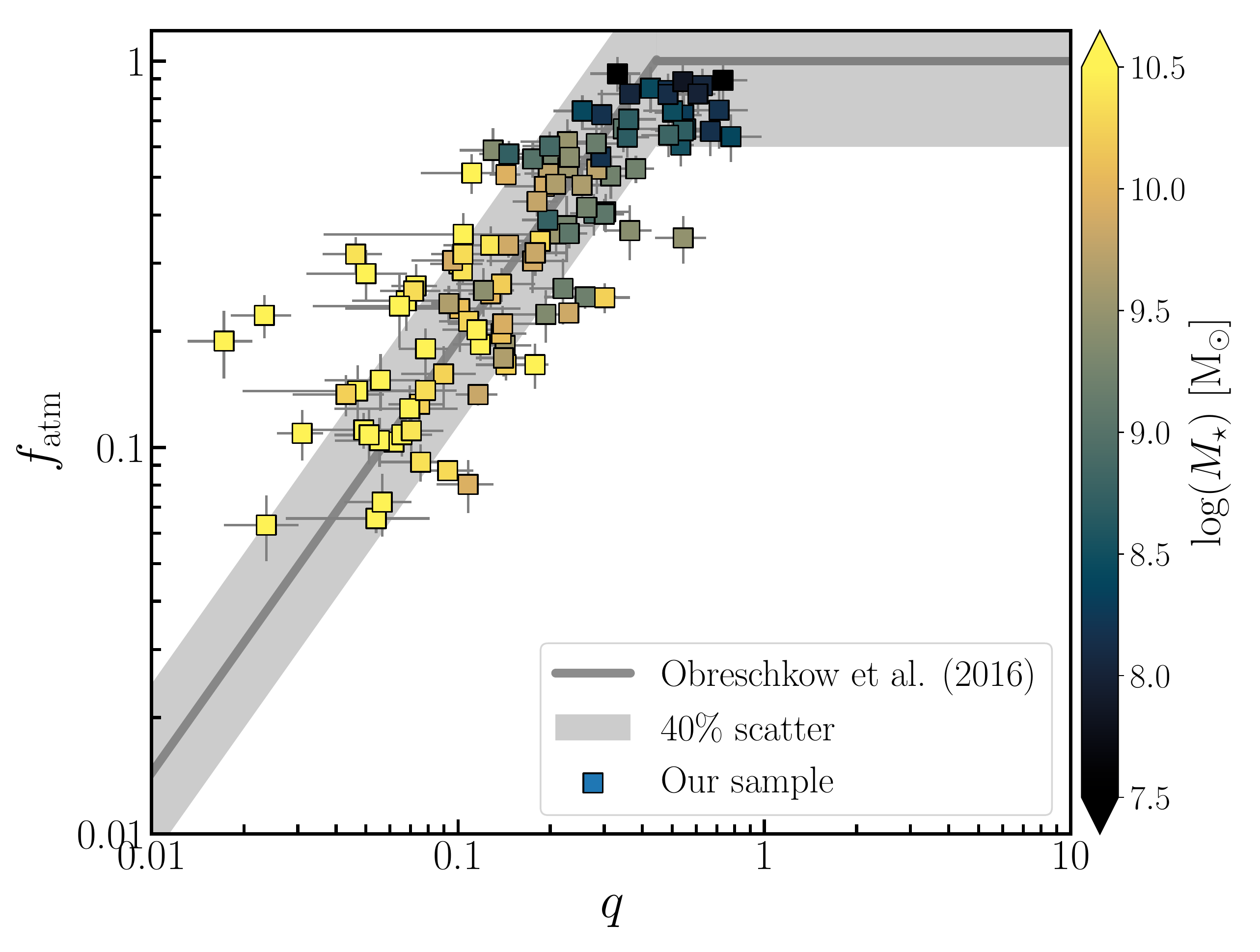}
    \caption{Plot shows the $f_{atm} -q$ relation with the stellar mass of the galaxies shown in the colour bar. A clear trend between $f_{atm}$ and $M_{\star}$ is observed, which is in fact the more commonly used $M_{\textrm{HI}}/M_{\star} - M_{\star}$ scaling relation. This empirical relation between $M_{\textrm{HI}}/M_{\star}$ and $M_{\star}$ can be explained by the ability of galaxies with a given disc stability to retain a certain fraction of \h1 gas as predicted by the models in \citetalias{obreschkow16}}
    \label{fig8}
\end{figure}

Fig.~\ref{fig8} shows the $f_{atm} - q - M_{\star}$ plane, with the stellar masses indicated in the colour bar. A smooth trend in stellar mass is observed, wherein low stellar mass systems in the sample contain higher gas fractions and vice versa. This observed trend between $f_{atm}$ and $M_{\star}$ is simply the more commonly used $M_{\textrm{HI}}/M_{\star} - M_{\star}$ scaling relation (see Fig.~\ref{fig2}), which can in fact be explained within the framework of the $f_{atm} - q$ model. Galaxies with larger $q$ are more stable and so can retain a larger fraction of \h1 and vice versa. This is explained by the dependence of $q$ on the total baryonic mass of galaxies i.e. $q \propto j_b/M_b \sim M_b^{-1/3}$, where $j_b = kM_b^{2/3}$ and $k$ is related to the halo spin parameter $\lambda$ (Bullock et al.~\citeyear{bullock01}). This leads to a relation between $f_{atm}$ and $M_b$ (see Section 3.2 in \citetalias{obreschkow16}). Galaxies that are outliers (either \h1-excess or \h1-deficient for their stellar mass) on the traditional $M_{\textrm{HI}}/M_{\star} - M_{\star}$ scaling relation are in fact not outliers on the $f_{atm} - q$ plane (see Lutz et al.~\citeyear{lutz17};~\citeyear{lutz18}; Murugeshan et al.~\citeyear{murugeshan2019}). Galaxies with similar mass, but significantly varying \h1 gas fractions, are systems that have different specific angular momenta, which in turn is heavily dependent on their formation and merger histories. Thus the AM of galaxies partly drive the observed scatter in the $M_{\textrm{HI}}/M_{\star} - M_{\star}$ scaling relation. This observation indicates that the $f_{atm} - q$ plane is a more fundamental and physically motivated scaling relation compared to the more traditional \h1 scaling relations connecting \h1 gas fraction and mass. The $f_{atm} - q$ parameter-space therefore can be used as a diagnostic plot to define \h1 deficiencies in galaxies by measuring their offset in $f_{atm}$ for their given $q$ values. This idea is introduced in a novel work by Li et al.~\citeyear{Li20}.

\vspace{-0.3cm}
\subsection{Relationship with SFR and SFE}
\label{Sec:fatm-q-SFR/SFE}

\begin{figure}
    \hspace*{-0.5cm}
    \includegraphics[width=8.5cm,height=6.5cm]{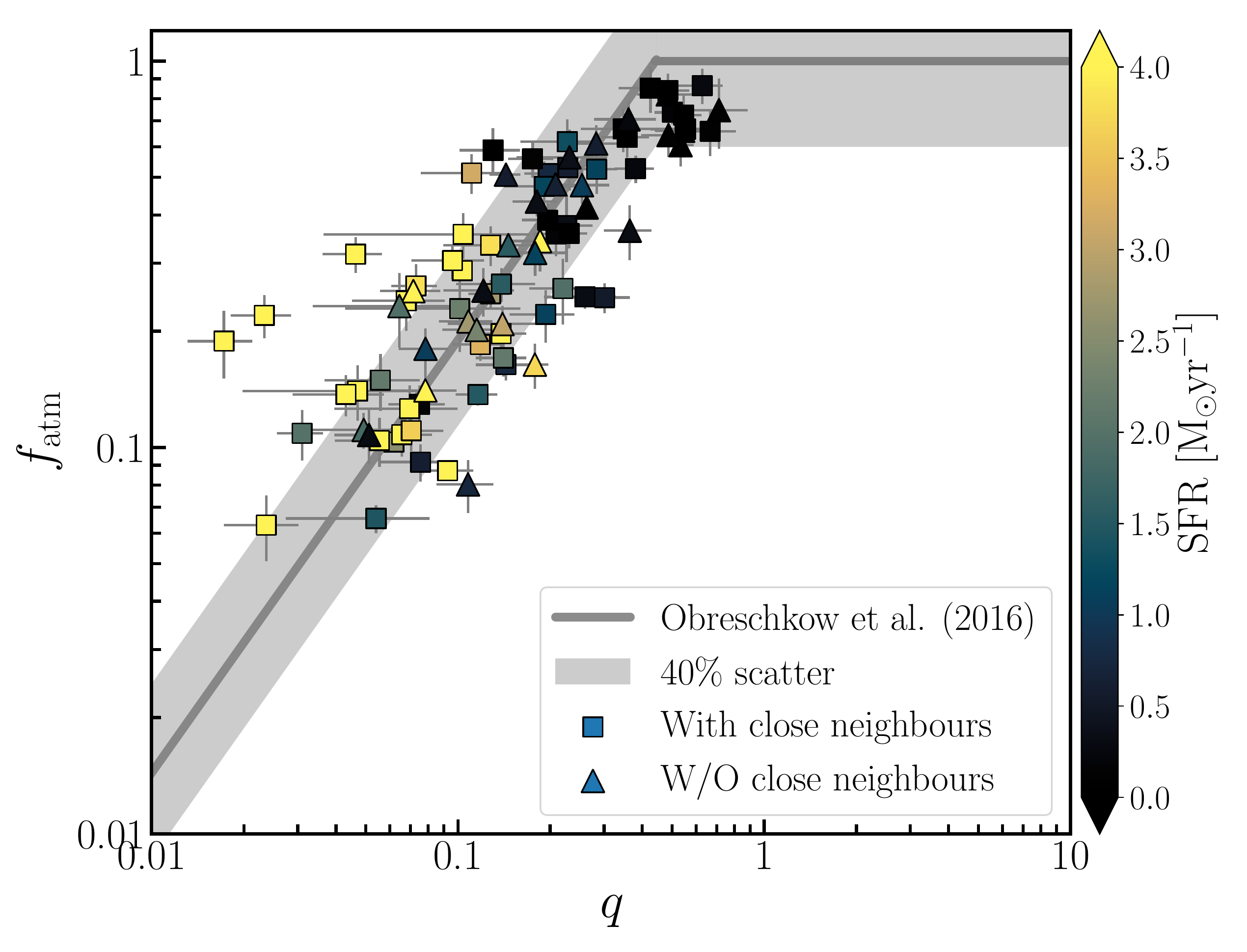}
    \caption{The $f_{atm} - q $ relation for a sub-sample of 86 WHISP galaxies for which SFR has been derived using $WISE$ W3 fluxes (Cluver et al.~\citeyear{Cluver17} and Jarrett et al.~\citeyear{Jarrett19}). A majority of galaxies with close neighbours (squares) and deviating from the relation, are seen to have elevated SFRs, compared to galaxies without close neighbours. This is in agreement with previous studies which find higher than average SFRs among galaxies that have close companions (Ellison et al.~\citeyear{ellison08}).}
    \label{fig9}
\end{figure}

In this section we present results pertaining to the star formation properties of our sample, such as their SFR and SFE. The SFR of the galaxies are derived from the $WISE$ (Wright et al.~\citeyear{Wright10}) W3 $12\mu$m flux using the relation from Cluver et al. (\citeyear{Cluver17}) and measurements as described in Jarrett et al. (\citeyear{Jarrett19}).Measured SFR is available for 86 of the original 114 galaxies in our sample. Fig.~\ref{fig9} shows the $f_{atm} - q - SFR$ relation for this sub-sample. We find that the majority of galaxies with close-neighbours which are deviating from the relation show elevated SFRs compared to galaxies without close neighbours. This supports our argument that galaxies that have interacted or are currently interacting, experience tidal forces that have the overall effect of lowering their $q$ values due to a net reduction in $j_b$ (see Barnes \& Hernquist~\citeyear{barnes96}; Cox et al.~\citeyear{cox08}; Hopkins \& Quataert~\citeyear{hopkins10}; Ellison et al.~\citeyear{ellison11} and references therein). A reduced $q$ value sets instabilities that funnel gas to the center leading to enhanced star formation. This naturally moves interacting galaxies to the left of the relation as observed in our case. Ellison et al. (\citeyear{ellison10}) study the influence of the local and intermediate environments on the SFR of galaxy pairs. They find that increased SFR is observed among galaxy pairs residing in low density environments. This is likely due to the fact that galaxies in low-density environments tend to typically have higher gas fractions. Our results fit well under this scenario, since we observe interacting pairs that are in low-density environments (based on their $\Sigma_2$ values; see Section~\ref{Sec:fatm-q-environment}) but are outliers on the $f_{atm} - q$ relation, showing enhanced SFR.

\begin{figure}
    \hspace*{-0.5cm}
    \includegraphics[width=8.5cm,height=6.5cm]{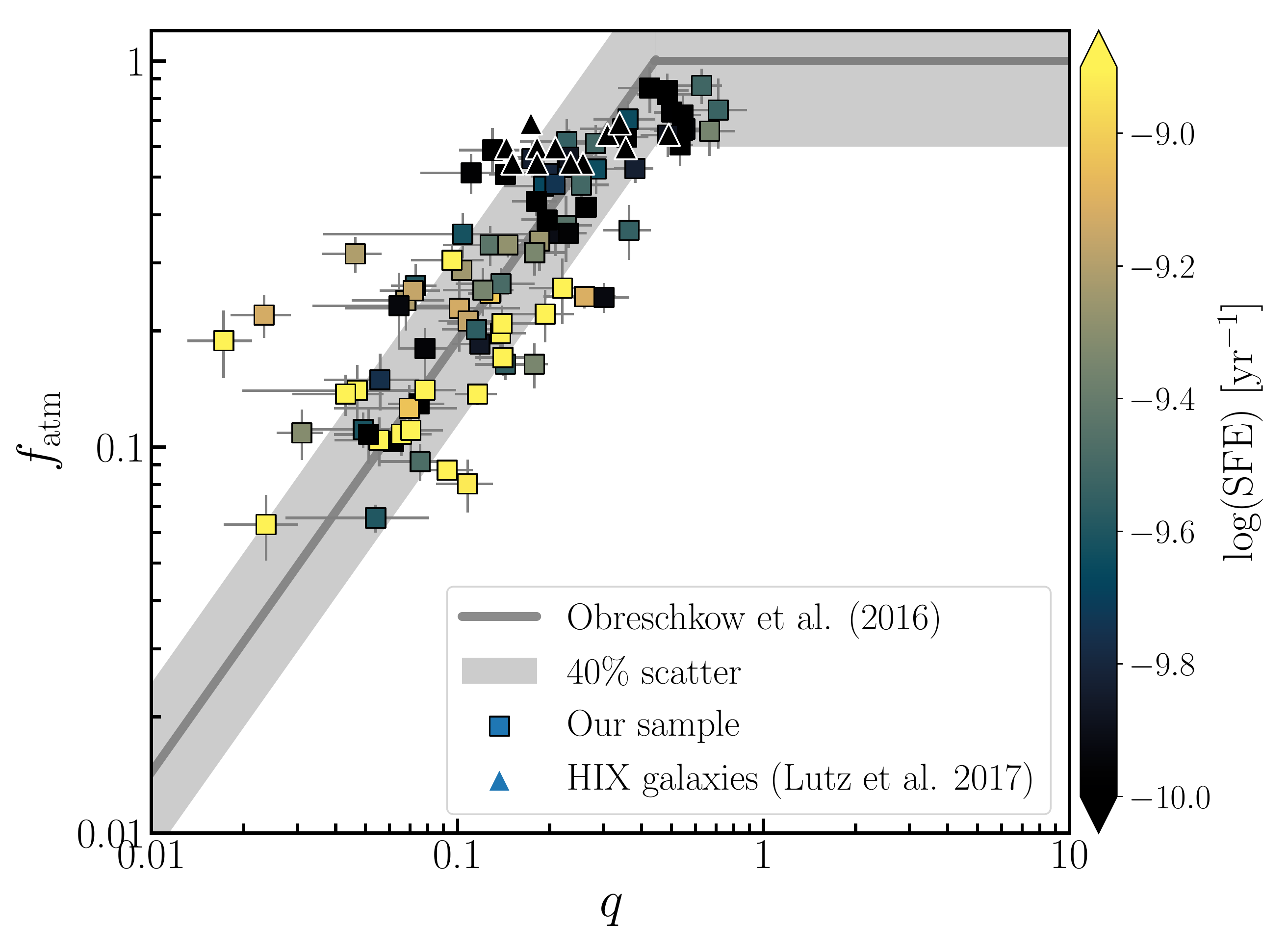}
    \caption{The $f_{atm} - q -$ SFE plane for a sub-sample 86 WHISP galaxies. Observations nicely agree with the theoretical models, which predict that galaxies with higher stability values will have a lower SFE and vice a versa. Also plotted are the \h1X sample (Lutz et al.~\citeyear{lutz17}) of \h1-excess galaxies. These galaxies are also observed to have low SFE owing to their high sAM.}
    \label{fig10}
\end{figure}

In terms of the SFE of the galaxies, Fig.~\ref{fig10} shows the trend for the sample. Overall, we observe that galaxies with higher $q$ value have a lower SFE and vice versa. This observation agrees well with the predictions of the original model proposed by \citetalias{obreschkow16}. Galaxies with a higher atomic disc stability prevent the \h1 gas from collapsing to form stars and hence have a low SFE, as opposed to galaxies with a lower stability which will show higher SFE. In addition to our sample, we have also included the 12 \h1X galaxies (Lutz et al.~\citeyear{lutz17}), which are some of the most \h1-rich galaxies in the local universe, and show that indeed they have very low SFE owing to their high sAM. Thus we show evidence of the influence of the atomic stability of discs on the SFE of galaxies, for a large sample spanning more than four decades in stellar mass.

\vspace{-0.3cm}
\subsection{Relationship with B/T ratio}
\label{Sec:fatm-q-B/T}

\begin{figure}
    \hspace*{-0.5cm}
    \includegraphics[width=8.5cm,height=6.5cm]{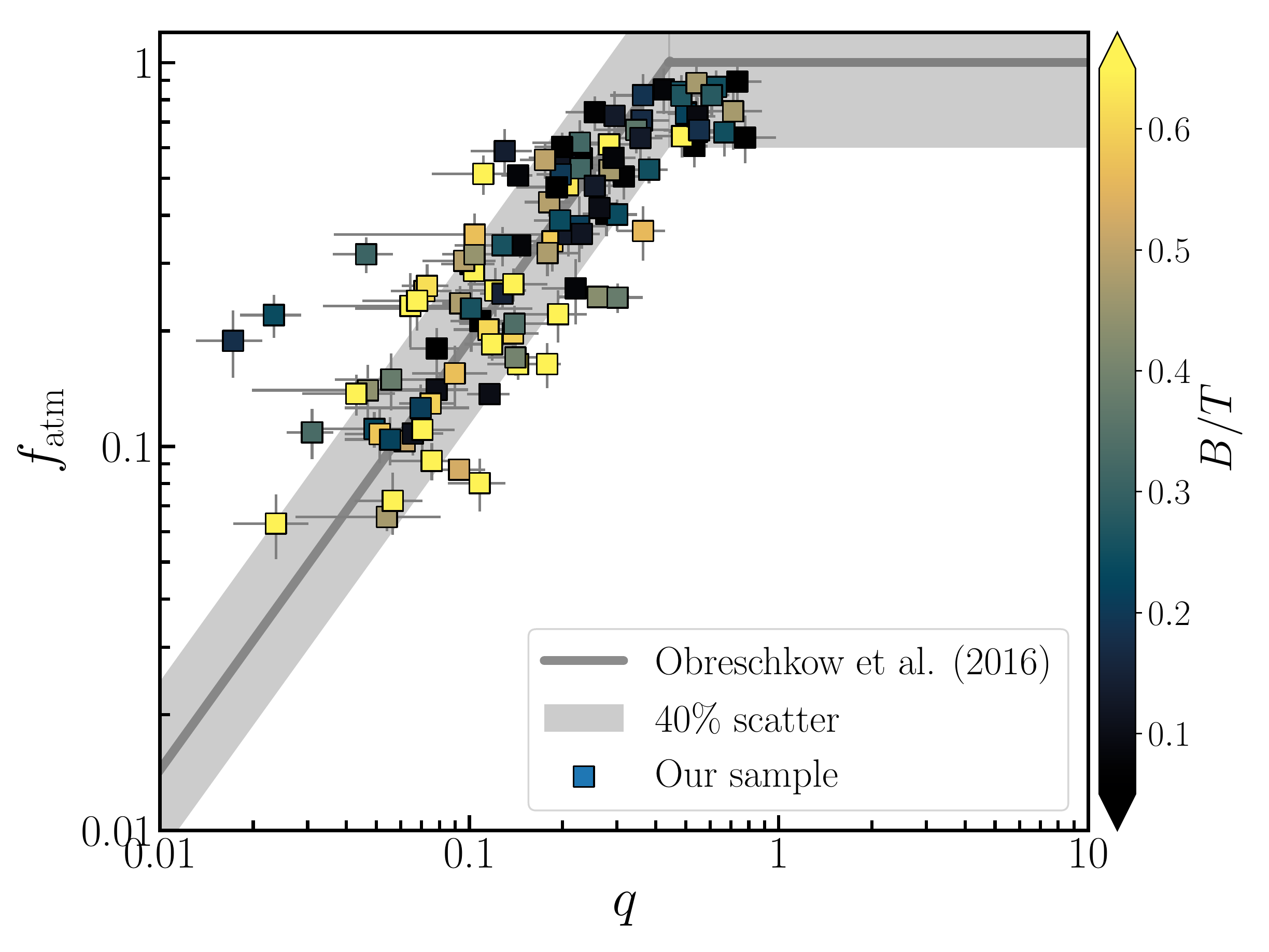}
    \caption{The $f_{atm} - q $ relation for a sub-sample of 106 WHISP galaxies for which B/T values have been computed using $WISE$ mosaics (Jarrett et al.~\citeyear{Jarrett19}). Low-mass galaxies with higher gas fractions are typically seen to have lower B/T values as opposed to the more evolved galaxies towards the bottom of the plot, which, depending on their merger histories evolve to become more bulgy due to loss of AM and build-up of mass over time. This presents a scenario where the influence of both mass and angular momentum on the gas fraction and the morphology of galaxies is evidenced.}
    \label{fig11}
\end{figure}

In this section we present the bulge-to-total (B/T) ratios of the galaxies in the sample and discuss their trend on the $f_{atm} - q$ plane. The B/T values for the sample have been derived using $WISE$ W1 3.4$\mu$m mosaics following the methods described in Jarrett et al. (\citeyear{Jarrett19}). The use of the W1 mosaics enable us to trace evolved stars without being affected by dust. The galaxies are fit with a series of elliptical annuli, and their radial surface brightness is determined. The derived surface brightness profile is then fit with a double S\'{e}rsic profile consisting of a bulge and disk component. Robust fits were derived for 106 of the original sample of 114 galaxies. Fig.~\ref{fig11} shows the $f_{atm} - q$ relation along with the B/T values for this sub-sample. We observe that, overall, bulge-dominated galaxies have a low \h1 gas fraction (low $f_{atm}$) and $q$ value. Galaxies with lower atomic stability are expected to be more bulge-dominated. This may be driven by both an increase in mass, which decreases the disc stability, and/or loss of specific AM, which makes the gas funnel to their centers, leading to central star formation. On the other hand, galaxies with a higher $q$ value are likely to be low-mass and/or high-spin systems, which resist the in-fall of gas and likely to be more disky. Our study shows for the first time how disk- and bulge-dominated galaxies are distributed on the $f_{atm} - q$ plane. This is an important secondary result showing the potential of this parameter-space in predicting not only the gas fraction for a given atomic disc stability, but also the morphology of galaxies.

\vspace{-0.4cm}
\section{Summary}
\label{Sec:conclusions}
In this work, we have studied the angular momentum and \h1 properties of galaxies as a function of their environment, as well as other global properties such as star formation and morphology. We have shown that AM is an integral property of galaxies and an important driver of their evolution. We have expanded on the original work of \citetalias{obreschkow16} on the $f_{atm} - q$ relation, and examined the influence of AM and environment on late-type galaxies. We summarise the main results from this study below:

\begin{itemize}
    \item We have measured precise specific baryonic angular momentum ($j_b$) and baryonic mass ($M_b$) for a sample of 114 WHISP galaxies. We find an unbroken power-law of the form $j_b \propto M_{b}^{0.55 \pm 0.02}$ over four orders of magnitude in $M_b$. 
    \item The sample was further divided into two sub-samples -- those with close neighbours and those without. We studied the behaviour of both sub-samples on the $f_{atm} - q$ plane, and find that galaxies without close neighbours follow the model prediction very tightly, with an intrinsic scatter of only $0.13$ dex, while the sub-sample of galaxies with close neighbours show a much larger scatter of $0.22$ dex. We attribute this larger scatter to the effects of past or ongoing interactions on the $j_b$ values of galaxies. 
    \item For the current sample, it is only their most local environment (close neighbours) that affect their location on the $f_{atm} - q$ relation, while their intermediate environment (group regime) does not appear to affect their $f_{atm}$, $j_b$ and $q$ values.
    \item Galaxies with close neighbours and deviating from the $f_{atm} - q$ relation exhibit enhanced star formation rates (SFR) compared to galaxies without close neighbours. This highlights the effects of external tidal perturbations on the disc stability of galaxies. Tidal fields tend to exert additional external torques on the \h1 gas, which makes the gas lose its AM and funnel to the centre to form stars, consequently enhancing the SFR.
    \item We use the bulge-to-total ratio (B/T) as a proxy for morphology and examine if the B/T values of the sample galaxies correlate on the $f_{atm} - q$ plane. We find evidence that, overall, galaxies with lower $q$ and $f_{atm}$ values tend to have higher B/T values. This result brings to light the importance of angular momentum and mass in determining not only the stability of discs, but also their influence on the \h1 gas fractions and morphology of galaxies. We show for the first time, a relation between atomic gas fraction, disc stability and the morphology of galaxies on the $f_{atm} - q$ plane.
\end{itemize}

The $f_{atm} - q$ plane is an important diagnostic plot that links the atomic gas fraction and disc stability to the mass, angular momentum, star formation and morphological properties of galaxies. Previous studies have shown the importance of this parameter-space in understanding how \h1-excess and \h1-deficient galaxies -- both outliers on the $M_\textrm{\h1}/M_{\star} - M_{\star}$ scaling relation, in fact, follow the $f_{atm} - q$ relation consistently (Lutz et al.~\citeyear{lutz17},~\citeyear{lutz18}; Murugeshan et al.~\citeyear{murugeshan2019}). In addition, Li et al. (\citeyear{Li20}) define a novel and more physically motivated way to quantify \h1 deficiencies in galaxies based on their atomic gas fraction offsets ($\Delta f_q$) in the $f_{atm} - q$ plane. In this work, we have shown that tidally interacting galaxies tend to lose their angular momentum and acquire lower $q$ values, deviating to the left of the relation. Ram pressure stripped galaxies on the other hand have significantly lower $f_{atm}$ values for their disc stability ($q$) and are observed to lie below the relation (Li et al.~\citeyear{Li20}). The fact that galaxies affected by tidal interactions and ram pressure occupy very different spaces on the $f_{atm} - q$ plane will be particularly important in disentangling the effects of the two processes in high-density environments, where both processes are prevalent. With upcoming high-resolution \h1 surveys such as WALLABY (Koribalski et al.~\citeyear{Koribalski20}), which has the potential for homogeneous volume-limited observations of thousands of galaxies from low- to high-density environments, and (DINGO; Meyer et al.~\citeyear{Meyer09}) with the potential to probe higher redshifts, we will be able to fully probe this parameter-space to study how galaxies behave as a function of both environment and redshift.

\vspace{-0.5cm}
\section*{Acknowledgements}
\label{sec:acknowledgements}
We thank the anonymous referee for their comments, which improved the overall quality of the paper. CM is supported by the Swinburne University Postgraduate Award (SUPRA). CM would like to thank Robert D\v{z}u{d}\v{z}ar, Luca Cortese and Adam Stevens for the useful discussions.

DO is a recipient of an Australian Research Council Future Fellowship (FT190100083) funded by the Australian Government.

KG and DO acknowledge the support of the Australian Research Council through the Discovery Project DP160102235.

MEC is a recipient of an Australian Research Council Future Fellowship (FT170100273) funded by the Australian Government.

We would like to thank the xGASS team for making the data publicly available.

This publication makes use of data products from the Two Micron
All Sky Survey, which is a joint project of the University
of Massachusetts and the Infrared Processing and Analysis Center/
California Institute of Technology, funded by the National Aeronautics
and Space Administration and the National Science Foundation. 

This research has made use of the NASA/IPAC Extragalactic
Database (NED), which is operated by the Jet Propulsion Laboratory,
California Institute of Technology, under contract with the
National Aeronautics and Space Administration. 

This publication makes use of data products from the Wide-field Infrared Survey Explorer, which is a joint project of the University of California, Los Angeles, and the Jet Propulsion 
Laboratory/California Institute of Technology, funded by the National Aeronautics and Space Administration.

Parts of the results in this work make use of the colourmaps in the CMasher package (van der Velden~\citeyear{ellert20}).

\vspace{-0.6cm}
\section*{Data Availability}
\noindent All data underlying this article is available within the article and enlisted in Table \ref{tab:sample} in Appendix \ref{appendix:properties_table}
\vspace{-0.5cm}


\bibliographystyle{mnras}
\bibliography{ref}


\appendix
\section{Properties of the sample galaxies}
\label{appendix:properties_table}
\begin{landscape}
\begin{table}
\caption{The galaxy sample. D is the galactocentric distance, V$_{sys}$ is the systemic velocity, $i$ is the inclination angle of the galaxy, $\sigma_{\textrm{\h1}}$ is the median of the \h1 dispersion velocity \\ 
computed from the 3D fit. The SFR values are borrowed from Jarrett et al.~\citeyear{Jarrett19}, which have been estimated using $WISE$ W3 fluxes. Bulge-to-total (B/T) values are derived using $WISE$ W1 3.4$\mu$m mosaics.}
\label{tab:sample}
\begin{tabular}{l l l c c c c c c c c c c c c c c}
\hline \hline
Name & RA & DEC & D & V$_{sys}$ & $i$ & V$_{max}$ & $\sigma_{\textrm{\h1}}$ & $\log(M_{\star})$ & $\log(M_{\textrm{\h1}})$ & $\log(M_{H_2})$ & $\log(M_b)$ & $j_b$ & $f_{atm}$ & $q$ & $\log(SFR)$ & B/T \\
    & [J2000]  & [J2000]  & Mpc  & \kms  & deg  & \kms  & \kms  & [M$_{\odot}$]  & [M$_{\odot}$]  & [M$_{\odot}$]  & [M$_{\odot}$]  & kpc~\kms  &   &   & [M$_{\odot}$ yr$^{-1}$] & \\
\hline
UGC89 	&	 00h09m53.41s 	&	 25d55m25.6s 	&	62.6	&	4569.3	&	58.9	&	201	&	11.5	&	11.09	&	9.81	&	9.74	&	11.14	&	1222.22	&	0.06	&	0.02	&	1.03	&	0.76	\\
UGC94 	&	 00h10m25.9s 	&	 25d49m55s 	&	62.9	&	4595.1	&	42.3	&	209	&	11.2	&	10.33	&	9.9	&	9.13	&	10.53	&	1359.59	&	0.32	&	0.1	&	 n/a 	&	0.45	\\
UGC232 	&	 00h24m38.69s 	&	 33d15m22.2s 	&	66.4	&	4845.3	&	47	&	145.9	&	13.8	&	10.63	&	9.93	&	9.36	&	10.76	&	2064.53	&	0.2	&	0.12	&	0.37	&	0.61	\\
UGC624 	&	 01h00m36.41s 	&	 30d40m08.3s 	&	65.5	&	4781.4	&	62.8	&	270.1	&	29.4	&	10.9	&	10.3	&	9.66	&	11.06	&	1905.7	&	0.24	&	0.11	&	1.1	&	0.89	\\
UGC731 	&	 01h10m43.98s 	&	 49d36m07.9s 	&	8.8	&	639.3	&	61.4	&	73.6	&	9.1	&	7.85	&	8.88	&	7.66	&	9.06	&	296.69	&	0.88	&	0.54	&	 n/a 	&	0.47	\\
UGC1541 	&	 02h03m27.94s 	&	 38d07m01.0s 	&	77.5	&	5658	&	55	&	228	&	16.4	&	10.95	&	10.09	&	9.65	&	11.05	&	2096.56	&	0.15	&	0.07	&	0.33	&	0.38	\\
UGC1550 	&	 02h03m44.8s 	&	 38d15m31s 	&	79	&	5768.1	&	75	&	225.1	&	21.4	&	11.03	&	10.53	&	9.81	&	11.21	&	2872.92	&	0.28	&	0.09	&	 n/a 	&	n/a	\\
UGC1886 	&	 02h26m00.49s 	&	 39d28m15.3s 	&	66.7	&	4868.4	&	59.6	&	263.9	&	8.8	&	10.84	&	10.22	&	9.59	&	10.99	&	3056.68	&	0.23	&	0.06	&	0.29	&	0.84	\\
UGC1913 	&	 02h27m16.88s 	&	 33d34m45.0s 	&	7.6	&	553.5	&	60.4	&	117.7	&	11.7	&	9.56	&	9.54	&	8.55	&	9.94	&	743.28	&	0.53	&	0.23	&	-0.21	&	0.32	\\
UGC2080 	&	 02h36m27.88s 	&	 38d58m11.7s 	&	12.4	&	903.6	&	23.3	&	145	&	5.3	&	9.75	&	9.69	&	8.71	&	10.11	&	1400.25	&	0.51	&	0.13	&	-0.12	&	0.2	\\
UGC2193 	&	 02h43m30.00s 	&	 37d20m28.8s 	&	7.1	&	518.4	&	26.5	&	 n/a 	&	3.1	&	9.21	&	8.62	&	7.96	&	9.36	&	347.01	&	0.25	&	0.13	&	-0.49	&	0.43	\\
UGC2487 	&	 03h01m42.37s 	&	 35d12m20.7s 	&	67.8	&	4948.5	&	37	&	318.9	&	10.6	&	11.47	&	10.43	&	10.14	&	11.54	&	8719.62	&	0.1	&	0.06	&	0.41	&	0.51	\\
UGC2503 	&	 03h03m34.75s 	&	 46d23m10.9s 	&	32.7	&	2388.6	&	63.3	&	252.6	&	7.6	&	10.88	&	9.87	&	9.56	&	10.96	&	2527.37	&	0.11	&	0.05	&	0.26	&	0.23	\\
UGC2800 	&	 03h40m02.46s 	&	 71d24m21.1s 	&	16.1	&	1176.9	&	65.8	&	113.3	&	8.1	&	9.23	&	9.16	&	8.19	&	9.59	&	649.48	&	0.5	&	0.32	&	 n/a 	&	0.07	\\
UGC2855 	&	 03h48m20.73s 	&	 70d07m58.4s 	&	16.5	&	1200.9	&	67	&	217.3	&	17.6	&	10.55	&	9.53	&	9.23	&	10.63	&	861.32	&	0.11	&	0.08	&	0.74	&	0.11	\\
UGC2916 	&	 04h02m33.86s 	&	 71d42m21.2s 	&	61.9	&	4520.1	&	50.6	&	198	&	13	&	10.72	&	10.17	&	9.48	&	10.88	&	1839.52	&	0.26	&	0.07	&	0.58	&	0.62	\\
UGC3013 	&	 04h23m27.10s 	&	 75d17m44.1s 	&	33.7	&	2462.7	&	41.9	&	211	&	10.9	&	10.73	&	9.92	&	9.43	&	10.83	&	1754.85	&	0.16	&	0.18	&	0.57	&	0.7	\\
UGC3205 	&	 04h56m14.88s 	&	 30d03m08.5s 	&	49.2	&	3591.6	&	66.9	&	211.7	&	11.2	&	10.74	&	9.98	&	9.46	&	10.86	&	2179.05	&	0.18	&	0.08	&	0.03	&	0.05	\\
UGC3354 	&	 05h47m18.22s 	&	 56d06m44.5s 	&	42.3	&	3087	&	69.3	&	185.9	&	18.7	&	10.33	&	9.95	&	9.15	&	10.55	&	1656.16	&	0.34	&	0.2	&	0.7	&	0.59	\\
UGC3371 	&	 05h56m38.60s 	&	 75d18m58.0s 	&	11.2	&	816.6	&	50.7	&	81.1	&	6.9	&	8.47	&	8.91	&	7.77	&	9.17	&	462.21	&	0.74	&	0.5	&	 n/a 	&	0.07	\\
UGC3382 	&	 05h59m47.73s 	&	 62d09m28.8s 	&	61.6	&	4499.1	&	40.8	&	146.1	&	6.4	&	10.63	&	9.61	&	9.31	&	10.7	&	739.85	&	0.11	&	0.02	&	0.3	&	0.33	\\
UGC3384 	&	 06h01m37.00s 	&	 73d07m00.0s 	&	14.9	&	1089.9	&	41.8	&	46.4	&	6	&	9.67	&	9.06	&	8.42	&	9.82	&	262.57	&	0.24	&	0.06	&	 n/a 	&	0.48	\\
UGC3546 	&	 06h50m08.66s 	&	 60d50m44.9s 	&	25.2	&	1841.4	&	54.5	&	202.4	&	10.7	&	10.4	&	9.39	&	9.08	&	10.48	&	857.69	&	0.11	&	0.07	&	0.56	&	0.7	\\
UGC3574 	&	 06h53m10.44s 	&	 57d10m40.0s 	&	19.8	&	1442.1	&	27.5	&	133.6	&	8.3	&	9.79	&	9.59	&	8.68	&	10.08	&	1130.3	&	0.43	&	0.18	&	-0.48	&	0.49	\\
UGC3580 	&	 06h55m30.86s 	&	 69d33m47.0s 	&	16.5	&	1201.8	&	65.8	&	119.1	&	7.6	&	9.16	&	9.29	&	8.23	&	9.63	&	691.31	&	0.61	&	0.28	&	-0.22	&	0.68	\\
UGC3642 	&	 07h04m20.30s 	&	 64d01m13.0s 	&	61.7	&	4501.2	&	39	&	317.6	&	16.6	&	11.13	&	10.39	&	9.85	&	11.25	&	5457.15	&	0.18	&	0.12	&	0.52	&	0.76	\\
UGC3711 	&	 07h10m13.58s 	&	 44d27m26.3s 	&	6	&	436.2	&	46.4	&	93	&	11.7	&	8.16	&	8.39	&	7.31	&	8.7	&	123.63	&	0.66	&	0.67	&	-0.96	&	0.25	\\
UGC3734 	&	 07h12m28.66s 	&	 47d10m00.1s 	&	13.4	&	974.7	&	31.7	&	130.1	&	6.3	&	9.43	&	8.86	&	8.19	&	9.59	&	320.8	&	0.25	&	0.12	&	-0.49	&	0.67	\\
UGC3993 	&	 07h55m43.97s 	&	 84d55m35.2s 	&	59.8	&	4368	&	34.2	&	152.9	&	6.7	&	10.72	&	9.7	&	9.4	&	10.79	&	2058.37	&	0.11	&	0.05	&	-0.52	&	0.58	\\
UGC4256 	&	 08h10m15.18s 	&	 33d57m23.9s 	&	72	&	5255.7	&	47	&	97.9	&	8.9	&	10.68	&	9.95	&	9.41	&	10.81	&	534.57	&	0.19	&	0.02	&	1.18	&	0.18	\\
UGC4273 	&	 08h12m57.92s 	&	 36d15m16.7s 	&	33.9	&	2472.6	&	66.2	&	158.2	&	7.5	&	10.1	&	9.48	&	8.85	&	10.25	&	1020.02	&	0.23	&	0.1	&	0.35	&	0.26	\\
UGC4284 	&	 08h14m40.12s 	&	 49d03m42.2s 	&	7.5	&	548.4	&	57	&	103.2	&	11.3	&	8.8	&	9.21	&	8.08	&	9.48	&	572.2	&	0.74	&	0.5	&	-0.77	&	0.22	\\
UGC4499 	&	 08h37m41.48s 	&	 51d39m08.6s 	&	9.5	&	691.5	&	50	&	74.2	&	7.1	&	8.05	&	8.74	&	7.56	&	8.95	&	198.7	&	0.82	&	0.36	&	 n/a 	&	0.19	\\
UGC4543 	&	 08h43m21.8s 	&	 45d44m10s 	&	26.9	&	1961.4	&	61	&	50.9	&	7.5	&	9.37	&	9.46	&	8.42	&	9.82	&	492.97	&	0.59	&	0.13	&	-1.41	&	0.14	\\
UGC4605 	&	 08h49m11.87s 	&	 60d13m16.0s 	&	18.5	&	1347.9	&	78	&	190.9	&	19	&	10.25	&	9.67	&	9.01	&	10.41	&	1909.47	&	0.24	&	0.33	&	-0.26	&	0.38	\\
UGC4806 	&	 09h09m33.71s 	&	 33d07m24.7s 	&	26.7	&	1948.2	&	69.7	&	173.4	&	13.8	&	9.88	&	9.75	&	8.8	&	10.2	&	947.49	&	0.47	&	0.19	&	0.08	&	0.02	\\
UGC4838 	&	 09h12m14.51s 	&	 44d57m17.4s 	&	36	&	2627.7	&	38.9	&	112.5	&	9	&	10.38	&	9.95	&	9.18	&	10.58	&	835.94	&	0.32	&	0.05	&	0.74	&	0.31	\\
UGC5079 	&	 09h32m10.11s 	&	 21d30m03.0s 	&	7.5	&	550.2	&	65	&	198.2	&	9	&	10.31	&	9.19	&	8.98	&	10.38	&	1053.21	&	0.09	&	0.09	&	0.67	&	0.53	\\
\hline	
\end{tabular}
\end{table}
\end{landscape}

\begin{landscape}
\begin{table}
\contcaption{}
\begin{tabular}{l l l c c c c c c c c c c c c c c}
\hline \hline
Name & RA & DEC & D & V$_{sys}$ & $i$ & V$_{max}$ & $\sigma_{\textrm{\h1}}$ & $\log(M_{\star})$ & $\log(M_{\textrm{\h1}})$ & $\log(M_{H_2})$ & $\log(M_b)$ & $j_b$ & $f_{atm}$ & $q$ & $\log(SFR)$ & B/T\\
    & [J2000]  & [J2000]  & Mpc  & \kms  & deg  & \kms  & \kms  & [M$_{\odot}$]  & [M$_{\odot}$]  & [M$_{\odot}$]  & [M$_{\odot}$]  & kpc~\kms  &   &   & [M$_{\odot}$ yr$^{-1}$] & \\
\hline
UGC5251 	&	 09h48m36.05s 	&	 33d25m17.4s 	&	20.3	&	1479	&	75.9	&	133.6	&	15.9	&	9.74	&	9.7	&	8.71	&	10.11	&	993.89	&	0.52	&	0.28	&	0.06	&	0.47	\\
UGC5253 	&	 09h50m22.23s 	&	 72d16m43.1s 	&	18.1	&	1323	&	47.7	&	 n/a 	&	11	&	10.38	&	9.88	&	9.16	&	10.56	&	1460.97	&	0.29	&	0.1	&	0.64	&	0.75	\\
UGC5316 	&	 09h55m40.6s 	&	 72d12m13s 	&	14.5	&	1058.7	&	65.8	&	109.2	&	6	&	9.48	&	9.11	&	8.3	&	9.7	&	1067.87	&	0.35	&	0.29	&	 n/a 	&	n/a	\\
UGC5414 	&	 10h03m57.35s 	&	 40d45m24.9s 	&	8.2	&	600.3	&	50.5	&	63.1	&	9.1	&	8.91	&	8.65	&	7.78	&	9.17	&	194.72	&	0.4	&	0.28	&	 n/a 	&	0.06	\\
UGC5557 	&	 10h18m16.86s 	&	 41d25m26.6s 	&	8.1	&	592.5	&	23	&	151.1	&	5.3	&	9.81	&	8.91	&	8.5	&	9.9	&	543.59	&	0.14	&	0.08	&	0.18	&	0.1	\\
UGC5589 	&	 10h21m47.59s 	&	 56d55m49.5s 	&	15.8	&	1151.7	&	51.9	&	108.2	&	8.6	&	9.53	&	9.18	&	8.36	&	9.76	&	598.6	&	0.36	&	0.21	&	-0.73	&	0.13	\\
UGC5685 	&	 10h29m19.94s 	&	 29d29m30.6s 	&	18.6	&	1356	&	74.3	&	209.4	&	8.7	&	9.82	&	9.4	&	8.63	&	10.03	&	945.34	&	0.32	&	0.18	&	0.06	&	0.48	\\
UGC5717 	&	 10h32m34.85s 	&	 65d02m27.9s 	&	23.1	&	1687.2	&	62.6	&	 n/a 	&	6.2	&	9.52	&	9.67	&	8.61	&	10	&	1596.5	&	0.62	&	0.23	&	0.12	&	0.32	\\
UGC5721 	&	 10h32m17.27s 	&	 27d40m07.6s 	&	7.4	&	537.3	&	63	&	82.1	&	9.9	&	8.17	&	8.93	&	7.74	&	9.14	&	290.44	&	0.84	&	0.48	&	-1.68	&	0.26	\\
UGC5786 	&	 10h38m45.86s 	&	 53d30m12.2s 	&	13.6	&	993.6	&	43.3	&	118.5	&	16.5	&	9.89	&	9.43	&	8.68	&	10.08	&	414.48	&	0.3	&	0.13	&	0.69	&	0.49	\\
UGC5789 	&	 10h39m09.46s 	&	 41d41m12.0s 	&	10.1	&	739.5	&	64	&	117.5	&	11.3	&	9.23	&	9.2	&	8.21	&	9.61	&	589.53	&	0.53	&	0.38	&	-0.63	&	0.25	\\
UGC5829 	&	 10h42m41.91s 	&	 34d26m56.0s 	&	8.6	&	629.4	&	52.5	&	45.2	&	8.7	&	8.86	&	8.97	&	7.93	&	9.32	&	208.78	&	0.6	&	0.2	&	 n/a 	&	0.06	\\
UGC5960 	&	 10h51m20.74s 	&	 32d45m59.0s 	&	8.8	&	645.3	&	66.6	&	81.4	&	16.5	&	8.18	&	8.62	&	7.48	&	8.88	&	139.33	&	0.75	&	0.71	&	-0.91	&	0.47	\\
UGC5997 	&	 10h53m54.86s 	&	 73d41m25.3s 	&	17.3	&	1263	&	69.3	&	145.5	&	10.1	&	9.65	&	9.53	&	8.59	&	9.98	&	1043.55	&	0.48	&	0.25	&	0.03	&	0.13	\\
UGC6128 	&	 11h04m02.9s 	&	 28d02m13s 	&	18.9	&	1377	&	41.7	&	138.3	&	11.1	&	9.56	&	8.81	&	8.28	&	9.68	&	260.99	&	0.18	&	0.14	&	 n/a 	&	0.27	\\
UGC6161 	&	 11h06m49.19s 	&	 43d43m23.7s 	&	10.4	&	756.6	&	55	&	76.4	&	9.1	&	8.13	&	8.81	&	7.63	&	9.03	&	246.46	&	0.82	&	0.48	&	-2.05	&	0.15	\\
UGC6225 	&	 11h11m30.97s 	&	 55d40m26.8s 	&	9.6	&	699.6	&	73.9	&	164.3	&	14.3	&	10.03	&	9.45	&	8.79	&	10.18	&	587.66	&	0.25	&	0.13	&	0.44	&	n/a	\\
UGC6263 	&	 11h14m10.89s 	&	 48d19m06.7s 	&	29.3	&	2137.5	&	54.4	&	 n/a 	&	7.6	&	10.59	&	9.7	&	9.29	&	10.68	&	1275.38	&	0.14	&	0.05	&	0.8	&	0.44	\\
UGC6283 	&	 11h15m52.01s 	&	 41d35m27.7s 	&	9.9	&	719.4	&	75.7	&	99.5	&	10.2	&	8.81	&	9.01	&	7.93	&	9.33	&	438.39	&	0.64	&	0.49	&	-0.88	&	0.72	\\
UGC6446 	&	 11h26m40.46s 	&	 53d44m48.0s 	&	8.8	&	645.9	&	56.7	&	75.6	&	8	&	9.07	&	8.8	&	7.93	&	9.33	&	347.56	&	0.4	&	0.3	&	 n/a 	&	0.24	\\
UGC6537 	&	 11h33m21.12s 	&	 47d01m45.1s 	&	11.9	&	866.1	&	53	&	164.5	&	9.1	&	9.86	&	9.47	&	8.67	&	10.07	&	814.16	&	0.33	&	0.15	&	0.19	&	0.08	\\
UGC6713 	&	 11h44m24.97s 	&	 48d50m06.7s 	&	12.3	&	899.7	&	43.5	&	63.1	&	5.5	&	8.29	&	8.67	&	7.55	&	8.94	&	258.65	&	0.73	&	0.38	&	-2.52	&	0.09	\\
UGC6778 	&	 11h48m38.19s 	&	 48d42m39.0s 	&	13.3	&	967.8	&	59	&	145.8	&	12	&	10.11	&	9.4	&	8.83	&	10.23	&	848.11	&	0.2	&	0.14	&	0.69	&	0.6	\\
UGC6786 	&	 11h49m09.46s 	&	 27d01m19.3s 	&	24.7	&	1799.7	&	68	&	219.6	&	10.4	&	10.22	&	9.68	&	8.98	&	10.38	&	1389.28	&	0.26	&	0.14	&	0.19	&	0.78	\\
UGC6787 	&	 11h49m15.37s 	&	 56d05m03.7s 	&	16.1	&	1176.9	&	63.9	&	246.6	&	13.6	&	10.25	&	9.44	&	8.95	&	10.35	&	1020.09	&	0.16	&	0.14	&	-0.12	&	0.73	\\
UGC6833 	&	 11h51m46.01s 	&	 38d00m54.4s 	&	12.6	&	919.5	&	54.7	&	92.2	&	5.8	&	9.24	&	9.01	&	8.12	&	9.52	&	440.18	&	0.42	&	0.18	&	-1.02	&	0.1	\\
UGC6840 	&	 11h52m07.01s 	&	 52d06m28.8s 	&	14.3	&	1046.7	&	56.6	&	90.7	&	10.4	&	9.05	&	9.3	&	8.2	&	9.6	&	570.07	&	0.67	&	0.35	&	-1.38	&	0.36	\\
UGC6884 	&	 11h54m58.71s 	&	 58d29m37.1s 	&	43.7	&	3190.2	&	52.2	&	71.4	&	7.6	&	10.49	&	9.84	&	9.23	&	10.63	&	558.84	&	0.22	&	0.02	&	0.72	&	0.24	\\
UGC6930 	&	 11h57m17.35s 	&	 49d16m59.1s 	&	10.7	&	777.6	&	31.1	&	111.7	&	8	&	8.05	&	8.94	&	7.74	&	9.13	&	463.04	&	0.86	&	0.63	&	-0.57	&	0.25	\\
UGC7030 	&	 12h03m09.61s 	&	 44d31m52.8s 	&	9.6	&	700.8	&	42.2	&	173.1	&	11.2	&	9.67	&	8.88	&	8.38	&	9.78	&	323.8	&	0.17	&	0.14	&	0.33	&	0.4	\\
UGC7075 	&	 12h05m22.71s 	&	 50d21m10.6s 	&	10.2	&	746.1	&	70.9	&	146.7	&	13.1	&	9.18	&	8.62	&	7.94	&	9.34	&	158.32	&	0.26	&	0.22	&	0.29	&	0.08	\\
UGC7081 	&	 12h05m34.2s 	&	 50d32m21s 	&	10.4	&	757.2	&	71.8	&	191.5	&	18	&	9.84	&	9.2	&	8.59	&	9.99	&	613	&	0.22	&	0.27	&	 n/a 	&	n/a	\\
UGC7095 	&	 12h06m08.45s 	&	 49d34m57.7s 	&	14.7	&	1075.2	&	73.3	&	185.4	&	10.2	&	9.92	&	9.25	&	8.66	&	10.06	&	671.49	&	0.21	&	0.14	&	0.48	&	0.34	\\
UGC7166 	&	 12h10m32.58s 	&	 39d24m20.6s 	&	13.6	&	995.7	&	33.6	&	102.1	&	8.4	&	10.24	&	9.34	&	8.93	&	10.33	&	472.85	&	0.14	&	0.04	&	0.86	&	0.85	\\
UGC7256 	&	 12h15m05.06s 	&	 33d11m50.4s 	&	14.9	&	1086.9	&	47	&	149.4	&	18.4	&	10.25	&	9.32	&	8.94	&	10.34	&	412.15	&	0.13	&	0.08	&	-0.73	&	0.6	\\
UGC7261 	&	 12h15m14.44s 	&	 20d39m30.9s 	&	11.9	&	870.6	&	38.7	&	58.7	&	7.5	&	8.99	&	9.02	&	8	&	9.4	&	252	&	0.56	&	0.18	&	-0.82	&	0.5	\\
UGC7323 	&	 12h17m30.18s 	&	 45d37m09.5s 	&	6.9	&	506.4	&	46	&	84.2	&	8.5	&	9.07	&	8.72	&	7.9	&	9.3	&	230.64	&	0.36	&	0.23	&	-1.24	&	0.12	\\
UGC7399 	&	 12h20m38.11s 	&	 46d17m30.0s 	&	7.1	&	520.2	&	57	&	91.1	&	7.9	&	8.44	&	8.69	&	7.6	&	9	&	298.34	&	0.67	&	0.55	&	-1.25	&	0.18	\\
\hline	
\end{tabular}
\end{table}
\end{landscape}

\begin{landscape}
\begin{table}
\contcaption{}
\begin{tabular}{l l l c c c c c c c c c c c c c c}
\hline \hline
Name & RA & DEC & D & V$_{sys}$ & $i$ & V$_{max}$ & $\sigma_{\textrm{\h1}}$ & $\log(M_*)$ & $\log(M_{\textrm{\h1}})$ & $\log(M_{H_2})$ & $\log(M_b)$ & $j_b$ & $f_{atm}$ & $q$ & $\log(SFR)$ & B/T\\
    & [J2000]  & [J2000]  & Mpc  & \kms  & deg  & \kms  & \kms  & [M$_{\odot}$]  & [M$_{\odot}$]  & [M$_{\odot}$]  & [M$_{\odot}$]  & kpc~\kms  &   &   & [M$_{\odot}$ yr$^{-1}$] & \\
\hline
UGC7483 	&	 12h24m11.17s 	&	 31d31m19.0s 	&	17.2	&	1254	&	78.7	&	103.1	&	12.2	&	9.4	&	9.06	&	8.24	&	9.63	&	551.4	&	0.36	&	0.36	&	-0.48	&	0.56	\\
UGC7559 	&	 12h27m05.15s 	&	 37d08m33.3s 	&	3	&	218.1	&	56.6	&	42	&	8.1	&	6.67	&	7.76	&	6.54	&	7.94	&	33.81	&	0.89	&	0.73	&	 n/a 	&	0.05	\\
UGC7603 	&	 12h28m44.11s 	&	 22d49m13.6s 	&	8.7	&	637.5	&	74.3	&	69.2	&	8	&	8.68	&	8.86	&	7.79	&	9.19	&	297.22	&	0.64	&	0.36	&	-1.85	&	0.13	\\
UGC7608 	&	 12h28m44.20s 	&	 43d13m26.9s 	&	7.4	&	538.5	&	32	&	54.7	&	6.2	&	7.08	&	8.65	&	7.42	&	8.82	&	151.81	&	0.93	&	0.33	&	 n/a 	&	n/a	\\
UGC7690 	&	 12h32m26.89s 	&	 42d42m14.8s 	&	7.4	&	537.3	&	48	&	47.7	&	6.8	&	8.74	&	8.46	&	7.6	&	9	&	122.96	&	0.39	&	0.2	&	-1.72	&	0.24	\\
UGC7766 	&	 12h35m57.65s 	&	 27d57m36.0s 	&	11.1	&	807.6	&	68	&	125.9	&	9.4	&	9.95	&	9.89	&	8.91	&	10.31	&	1346.12	&	0.51	&	0.14	&	-0.24	&	0.08	\\
UGC7861 	&	 12h41m52.7s 	&	41d16m26s 	&	8.5	&	621.3	&	42.2	&	45.6	&	5	&	8.73	&	8.79	&	7.76	&	9.16	&	153.19	&	0.57	&	0.12	&	 n/a 	&	n/a	\\
UGC7916 	&	 12h44m25.14s 	&	 34d23m11.5s 	&	8.3	&	607.5	&	67.2	&	39.9	&	9.3	&	8.17	&	8.56	&	7.44	&	8.83	&	92.51	&	0.73	&	0.29	&	 n/a 	&	0.13	\\
UGC7971 	&	 12h48m22.87s 	&	 51d09m52.9s 	&	6.4	&	468.3	&	37	&	42.6	&	7.2	&	8.18	&	8.22	&	7.2	&	8.6	&	69.22	&	0.56	&	0.29	&	 n/a 	&	0.08	\\
UGC7989 	&	 12h50m26.58s 	&	 25d30m02.9s 	&	16.5	&	1206.9	&	49.9	&	232.1	&	10.6	&	11.01	&	9.75	&	9.67	&	11.07	&	2579	&	0.07	&	0.05	&	0.16	&	0.47	\\
UGC8403 	&	 13h21m56.4s 	&	 38d44m05s 	&	13.4	&	975.6	&	51.9	&	125.3	&	11	&	9.38	&	9.13	&	8.25	&	9.65	&	525.06	&	0.41	&	0.3	&	 n/a 	&	n/a	\\
UGC8699 	&	 13h45m08.71s 	&	 41d30m12.2s 	&	34.6	&	2527.8	&	72.1	&	190.9	&	9.5	&	10.26	&	9.42	&	8.96	&	10.36	&	930.89	&	0.16	&	0.09	&	 n/a 	&	0.57	\\
UGC8700 	&	 13h45m19.2s 	&	 41d42m45s 	&	35.3	&	2574.9	&	75	&	247.1	&	14.4	&	10.68	&	9.47	&	9.34	&	10.74	&	924.83	&	0.07	&	0.06	&	 n/a 	&	0.87	\\
UGC8709 	&	 13h46m23.67s 	&	 43d52m20.4s 	&	33	&	2410.8	&	74	&	200.6	&	11.9	&	10.41	&	10.01	&	9.22	&	10.62	&	1926.58	&	0.33	&	0.13	&	0.58	&	0.26	\\
UGC8863 	&	 13h56m16.67s 	&	 47d14m08.5s 	&	24.6	&	1794.3	&	65	&	209.9	&	10	&	10.36	&	9.26	&	9.03	&	10.43	&	870.25	&	0.09	&	0.08	&	-0.22	&	0.73	\\
UGC9211 	&	 14h22m32.17s 	&	 45d23m01.9s 	&	9.4	&	686.4	&	57.3	&	57.8	&	5.7	&	8	&	8.69	&	7.51	&	8.91	&	216.6	&	0.82	&	0.36	&	 n/a 	&	0.28	\\
UGC9366 	&	 14h32m46.85s 	&	 49d27m28.4s 	&	29	&	2115.6	&	63.5	&	229.4	&	12.1	&	10.71	&	9.67	&	9.38	&	10.78	&	1191.17	&	0.1	&	0.06	&	1.04	&	0.22	\\
UGC9648 	&	 14h58m59.6s 	&	 53d55m24s 	&	46.3	&	3378.3	&	65.2	&	135.4	&	9.3	&	9.88	&	9.42	&	8.67	&	10.07	&	954.65	&	0.3	&	0.18	&	 n/a 	&	n/a	\\
UGC9753 	&	 15h09m46.73s 	&	 57d00m00.7s 	&	10.6	&	772.5	&	75.7	&	136.4	&	9.5	&	9.35	&	8.7	&	8.09	&	9.49	&	269.22	&	0.22	&	0.19	&	0.05	&	0.82	\\
UGC9797 	&	 15h15m23.32s 	&	 55d31m02.5s 	&	46.5	&	3392.4	&	49.6	&	198	&	10	&	10.7	&	10.35	&	9.53	&	10.93	&	3817.03	&	0.36	&	0.1	&	0.73	&	0.57	\\
UGC9969 	&	 15h39m37.09s 	&	 59d19m55.0s 	&	34.5	&	2518.8	&	63.7	&	286.7	&	7	&	10.8	&	9.86	&	9.49	&	10.89	&	3287.62	&	0.13	&	0.07	&	0.82	&	0.21	\\
UGC10310 	&	 16h16m18.35s 	&	 47d02m47.1s 	&	9.8	&	716.4	&	47.3	&	64.5	&	5.8	&	8.52	&	8.64	&	7.59	&	8.99	&	207.49	&	0.61	&	0.29	&	-1.62	&	0.06	\\
UGC10359 	&	 16h20m58.16s 	&	 65d23m26.0s 	&	12.5	&	910.5	&	49.3	&	129.8	&	6.5	&	9.66	&	9.54	&	8.59	&	9.99	&	1344.04	&	0.48	&	0.21	&	-0.2	&	0.63	\\
UGC10445 	&	 16h33m47.62s 	&	 28d59m05.2s 	&	13.2	&	963.6	&	56.8	&	65.6	&	8.8	&	8.69	&	9.03	&	7.91	&	9.31	&	358.7	&	0.71	&	0.36	&	-0.62	&	0.17	\\
UGC10470 	&	 16h32m39.20s 	&	 78d11m53.4s 	&	18.7	&	1362.9	&	46.4	&	119.8	&	8.2	&	10.33	&	9.76	&	9.09	&	10.49	&	1151.03	&	0.25	&	0.07	&	0.6	&	0.61	\\
UGC10502 	&	 16h37m37.69s 	&	 72d22m28.8s 	&	59	&	4310.1	&	32.7	&	206.1	&	5.5	&	10.4	&	9.93	&	9.19	&	10.59	&	1584.63	&	0.3	&	0.05	&	 n/a 	&	0.03	\\
UGC10564 	&	 16h46m21.99s 	&	 70d21m31.7s 	&	15.5	&	1129.8	&	71.2	&	84.1	&	10.3	&	8.44	&	9.26	&	8.06	&	9.46	&	516.03	&	0.85	&	0.43	&	-1.18	&	0.08	\\
UGC11124 	&	 18h07m27.50s 	&	 35d33m48.0s 	&	22.1	&	1614	&	49.5	&	81.7	&	9.2	&	9.29	&	9.28	&	8.28	&	9.68	&	511.68	&	0.54	&	0.23	&	 n/a 	&	0.02	\\
UGC11218 	&	 18h19m46.41s 	&	 74d34m06.1s 	&	20.3	&	1485	&	63	&	179.1	&	10.2	&	10.33	&	9.44	&	9.02	&	10.42	&	871.75	&	0.14	&	0.08	&	0.65	&	0.1	\\
UGC11670 	&	 21h03m33.58s 	&	 29d53m50.9s 	&	10.7	&	779.4	&	66.5	&	162.7	&	10.5	&	9.95	&	8.78	&	8.61	&	10.01	&	451.67	&	0.08	&	0.11	&	-0.14	&	0.86	\\
UGC11852 	&	 21h55m59.31s 	&	 27d53m54.3s 	&	80.2	&	5853.9	&	65	&	157.6	&	10	&	10.52	&	10.46	&	9.49	&	10.88	&	3643.8	&	0.51	&	0.11	&	0.5	&	0.7	\\
UGC11861 	&	 21h56m24.00s 	&	 73d15m38.6s 	&	20.3	&	1482	&	54	&	146.2	&	9.6	&	10.27	&	9.6	&	9	&	10.4	&	1225.59	&	0.21	&	0.11	&	0.44	&	0.05	\\
UGC11909 	&	 22h06m16.17s 	&	 47d15m04.4s 	&	15.1	&	1105.8	&	76	&	152.2	&	18.5	&	9.42	&	9.45	&	8.44	&	9.83	&	454.94	&	0.56	&	0.29	&	-0.43	&	0.04	\\
UGC11951 	&	 22h12m30.12s 	&	 45d19m42.5s 	&	14.8	&	1078.8	&	71.4	&	121.1	&	13	&	9.27	&	8.95	&	8.11	&	9.51	&	243.77	&	0.38	&	0.23	&	-0.49	&	0.21	\\
UGC12043 	&	 22h27m50.52s 	&	 29d05m45.5s 	&	13.8	&	1008.6	&	70.9	&	90.8	&	10.7	&	8.7	&	8.94	&	7.85	&	9.25	&	388.11	&	0.66	&	0.54	&	-1.22	&	0.08	\\
UGC12082 	&	 22h34m10.82s 	&	 32d51m37.8s 	&	11	&	802.5	&	48.4	&	51.8	&	6.5	&	8.42	&	8.85	&	7.72	&	9.11	&	219.28	&	0.74	&	0.25	&	 n/a 	&	0.07	\\
UGC12212 	&	 22h50m30.33s 	&	 29d08m18.4s 	&	12.1	&	886.5	&	46.1	&	106.9	&	7.9	&	8.4	&	8.58	&	7.51	&	8.91	&	343.58	&	0.64	&	0.78	&	 n/a 	&	0.06	\\
UGC12732 	&	 23h40m39.86s 	&	 26d14m11.1s 	&	10.2	&	747.6	&	36.3	&	91.5	&	6.6	&	9.26	&	9.25	&	8.25	&	9.65	&	595.33	&	0.54	&	0.2	&	 n/a 	&	0.12	\\
\hline
\end{tabular}
\end{table}
\end{landscape}

\bsp	
\label{lastpage}
\end{document}